\documentstyle[]{mn}
\newif\ifAMStwofonts

\ifoldfss
  \newcommand{\rmn}[1] {{\rm #1}}

  \ifCUPmtlplainloaded \else
    \NewTextAlphabet{textbfit} {cmbxti10} {}
    \NewTextAlphabet{textbfss} {cmssbx10} {}
    \NewMathAlphabet{mathbfit} {cmbxti10} {} 
    \NewMathAlphabet{mathbfss} {cmssbx10} {} 
  \fi
  \ifAMStwofonts
    \ifCUPmtlplainloaded \else
      \NewSymbolFont{upmath} {eurm10}
      \NewSymbolFont{AMSa} {msam10}
      \NewMathSymbol{\upi}     {0}{upmath}{19}
      \NewMathSymbol{\umu}     {0}{upmath}{16}
      \NewMathSymbol{\upartial}{0}{upmath}{40}
      \NewMathSymbol{\leqslant}{3}{AMSa}{36}
      \NewMathSymbol{\geqslant}{3}{AMSa}{3E}

      \let\leq=\leqslant 
      \let\geq=\geqslant 
    \fi
  \fi
\fi 

\ifnfssone
  \newmathalphabet{\mathit}
  \addtoversion{normal}{\mathit}{cmr}{m}{it}
  \addtoversion{bold}{\mathit}{cmr}{bx}{it}
  \newcommand{\rmn}[1] {\mathrm{#1}}

  \newmathalphabet{\mathbfit} 
  \addtoversion{normal}{\mathbfit}{cmr}{bx}{it}
  \addtoversion{bold}{\mathbfit}{cmr}{bx}{it}
  \newmathalphabet{\mathbfss} 
  \addtoversion{normal}{\mathbfss}{cmss}{bx}{n}
  \addtoversion{bold}{\mathbfss}{cmss}{bx}{n}
  \ifAMStwofonts
    \ifCUPmtlplainloaded \else
      \UseAMStwoboldmath
      \makeatletter
      \new@mathgroup\upmath@group
      \define@mathgroup\mv@normal\upmath@group{eur}{m}{n}
      \define@mathgroup\mv@bold\upmath@group{eur}{b}{n}
      \edef\UPM{\hexnumber\upmath@group}
      \new@mathgroup\amsa@group
      \define@mathgroup\mv@normal\amsa@group{msa}{m}{n}
      \define@mathgroup\mv@bold\amsa@group{msa}{m}{n}
      \edef\AMSa{\hexnumber\amsa@group}
      \makeatother
      \mathchardef\upi="0\UPM19
      \mathchardef\umu="0\UPM16
      \mathchardef\upartial="0\UPM40
      \mathchardef\leqslant="3\AMSa36
      \mathchardef\geqslant="3\AMSa3E

      \let\leq=\leqslant 
      \let\geq=\geqslant 
    \fi
  \fi
\fi 

\ifnfsstwo
  \newcommand{\rmn}[1] {\mathrm{#1}}

  \DeclareMathAlphabet{\mathbfit}{OT1}{cmr}{bx}{it}
  \SetMathAlphabet\mathbfit{bold}{OT1}{cmr}{bx}{it}
  \DeclareMathAlphabet{\mathbfss}{OT1}{cmss}{bx}{n}
  \SetMathAlphabet\mathbfss{bold}{OT1}{cmss}{bx}{n}
  \ifAMStwofonts
    \ifCUPmtlplainloaded \else
      \DeclareSymbolFont{UPM}{U}{eur}{m}{n}
      \SetSymbolFont{UPM}{bold}{U}{eur}{b}{n}
      \DeclareSymbolFont{AMSa}{U}{msa}{m}{n}
      \DeclareMathSymbol{\upi}{0}{UPM}{"19}
      \DeclareMathSymbol{\umu}{0}{UPM}{"16}
      \DeclareMathSymbol{\upartial}{0}{UPM}{"40}
      \DeclareMathSymbol{\leqslant}{3}{AMSa}{"36}
      \DeclareMathSymbol{\geqslant}{3}{AMSa}{"3E}

      \let\leq=\leqslant 
      \let\geq=\geqslant 
    \fi
  \fi
\fi 

\ifCUPmtlplainloaded \else
  \ifAMStwofonts \else 
    \def\upi{\pi}
    \def\umu{\mu}
    \def\upartial{\partial}
  \fi
\fi

\newcommand{\samethanks}
	{{\Huge $^\star$}}

\newcommand{\etal}
	{et al.}
\newcommand{\eg}
	{e.g.}
\newcommand{\cf}
	{c.f.}
\newcommand{\ie}
	{i.e.}
\newcommand{\eq}[1]
	{equation~(\ref{equation:#1})}
\newcommand{\eqs}[1]
	{equations~(\ref{equation:#1})}
\newcommand{\Eq}[1]
        {Equation~(\ref{equation:#1})}

\newcommand{\sect}[1]
	{Section~\ref{section:#1}}

\newcommand{\sects}[1]
        {Sections~\ref{section:#1}}
\newcommand{\Sect}[1]
        {Section~\ref{section:#1}}

\newcommand{\fig}[1]
	{Fig.~\ref{figure:#1}}
\newcommand{\figs}[1]
        {Figs.~\ref{figure:#1}}
\newcommand{\Fig}[1]
        {Fig.~\ref{figure:#1}}
\newcommand{\Figs}[1]
        {Figs.~\ref{figure:#1}}
\newlength{\singlefigureheight}
\newlength{\doublefigureheight}
\newlength{\triplefigureheight}
\newlength{\squarefigureheight}
\newcommand{\AaA}
        {A\&A}

\newcommand{\AJ}
        {AJ}
\newcommand{\ApJ}
        {ApJ}

\newcommand{\ARAA}
        {ARA\&A}

\newcommand{\MNRAS}
        {MNRAS}
\newcommand{\PhilTransA}
        {Phil.\ Trans.\ of the Royal Soc.\ A}

\newcommand{\PRD}
        {Phys.\ Rev.\ D}

\newcommand{\los}
        {line-of-sight}

\newcommand{\FL}
        {FL}

\begin{document}

\title[Lensing by elliptical galaxies]
{Gravitational lensing by elliptical galaxies}

\author[D.\ J.\ Mortlock and R.\ L.\ Webster]
       {
        Daniel J.\ Mortlock$^{1,2,3}$\thanks{
		E-mail: mortlock@ast.cam.ac.uk (DJM);
		rwebster@physics. unimelb.edu.au (RLW)}
	and Rachel L.\ Webster$^1$\samethanks\ \\
        $^1$School of Physics, The University of Melbourne, Parkville,
        Victoria 3052, Australia \\
        $^2$Astrophysics Group, Cavendish Laboratory, Madingley Road,
        Cambridge CB3 0HE, U.K. \\
        $^3$Institute of Astronomy, Madingley Road, Cambridge
        CB3 0HA, U.K. \\
       }

\date{
Accepted. 
Received; in original form 2000 April 19}

\pagerange{\pageref{firstpage}--\pageref{lastpage}}
\pubyear{2000}

\label{firstpage}

\maketitle

\begin{abstract}
The fraction of high-redshift sources which are multiply-imaged
by intervening galaxies is strongly dependent on the cosmological
constant, and so can be a useful probe of the cosmological
model. However its power is 
limited by various systematic (and random) uncertainties in the
calculation of lensing probabilities, one of the most important
of which is the dynamical normalisation of elliptical galaxies.
Assuming ellipticals' mass distributions can be modelled as 
isothermal spheres, the mass normalisation depends on: the 
velocity anisotropy; the luminosity density; the core radius;
and the area over which the velocity dispersion is measured. The 
differences in the lensing probability and optical depth produced
by using the correct normalisation can be comparable to the
differences between even the most extreme cosmological models.
The existing data is not sufficient to determine the correct
normalisation with enough certainty to allow lensing statistics
to be used to their full potential. However, as 
the correct lensing probability is almost certainly higher than is usually
assumed, upper bounds on the cosmological constant are 
not weakened by these possibilities.
\end{abstract}


\section{Introduction}
\label{section:intro_core}

The fraction of high-redshift quasars which are multiply-imaged
due to gravitational lensing
is determined mainly by the cosmological model and the population
of potential lenses. Many of the early investigations 
into the statistics of quasar lensing
(\eg\ Press \& Gunn 1973; Turner, Ostriker \& Gott 1984; 
Kochanek \& Blandford 1987; Fukugita \& Turner 1991; Maoz \& Rix 1993)
focussed on the deflector population, but 
more recent studies have emphasised the cosmological 
possibilities. Specifically, Turner (1990) and 
Fukugita, Futumase \& Kasai (1990) found that 
the lensing probability increases very rapidly with the 
(normalised) cosmological constant, $\Omega_{\Lambda_0}$, 
but depends only weakly on the (normalised) matter density, 
$\Omega_{{\rmn m}_0}$. 
One of the most stringent upper limits that can be placed on the
value of the cosmological constant
is due to the low number of lenses detected -- 
both Kochanek (1996b) and Falco, Kochanek \& Mu\~{n}oz (1998)
found that $\Omega_{\Lambda_0} \la 0.65$, with 95 per cent confidence.
These results are only marginally consistent with a number of 
independent cosmological measurements, such as high-redshift 
supernova observations (\eg\ Schmidt \etal\ 1998; Perlmutter \etal\ 1999)
and cosmic microwave background measurements (\eg\ 
Lineweaver 1998; Efstathiou \etal\ 1999),
which imply that 
$\Omega_{\Lambda_0} = 0.7 \pm 0.2$, again
at the 95 per cent confidence level.
Further, the low density implied by cluster observations 
(\eg\ Bahcall, Fan \& Cen 1997), 
combined with the inflationary requirement of 
a flat universe (\eg\ Guth 1981; 
Kolb \& Turner 1989) also imply a high value of $\Omega_{\Lambda_0}$.
It is thus very important to accurately assess both the random 
and systematic uncertainties on the lensing constraints.

Some of the random uncertainties in the lens statistics are being 
steadily reduced as new surveys better constrain 
the deflector and source populations and more lenses are discovered.
If ellipticals 
do completely dominate the lensing probability (\eg\ Turner \etal\ 
1984; Kochanek 1996b), any improvements in the knowledge of the
number density of galaxies must be accompanied by accurate type 
information. Both the 
the Sloan Digital Sky Survey (\eg\ Szalay 1998; Loveday \& Pier 1998)
and the 
2 degree Field galaxy redshift survey (\eg\ Colless 1999; Folkes \etal\ 1999)
should decrease the errors
on the type-specific luminosity functions by up to an order of magnitude. 
These two projects will also greatly reduce the uncertainties in the 
quasar luminosity function, as well as yielding more lensed quasars
than are known to date (\eg\ Loveday \& Pier 1998; Boyle \etal\ 1999a,b; 
Mortlock \& Webster 2000b).

However, comparable progress in reducing 
the various systematic uncertainties is 
unlikely to be as rapid or as certain. 
Firstly, 
dust in the lensing galaxies can obscure multiply-imaged quasars from 
lens surveys. 
There is some evidence that it is an unimportant 
effect (\eg\ Kochanek \etal\ 1999; Falco \etal\ 1999), but it 
is also possible that it dominates the statistics 
(Malhotra, Rhoads \& Turner 1997).
There are a number of studies of dust in local galaxies, but 
it is reddening in high-redshift galaxies that is more important
to lensing statistics.
The only measurements of obscuration in such galaxies 
comes from lensed quasars, as the colours of the various images
of the one source can be compared. Using this technique Falco \etal\ (1999)
found that ellipticals with redshifts of up to 
$\sim 1$ have minimal dust content (the difference in the extinctions 
between different lines-of-sight being
only $\Delta E (B - V) \simeq 0.2$ mag).

Another potential limitation on the accuracy of lens statistics 
is uncertainty in the mass evolution of galaxies. 
Keeton, Kochanek \& Falco (1998) and Kochanek \etal\ (2000) have
used lens galaxies to measure the fundamental plane 
(\eg\ Dressler \etal\ 1997) of field ellipticals at moderate
redshifts, but very little could be inferred about 
the mass evolution of the population.
Assuming the present-day population of ellipticals formed from the
mergers of spirals (or other smaller halos), the high-redshift 
deflector population should consist of a larger number 
of less massive objects. If the total mass in halos is conserved
the lensing optical depth is independent of the evolution,
but the average image separation is decreased (\eg\ Rix \etal\ 1994). 
Mao \& Kochanek (1994) used this fact to show that the
known quasar lenses were best explained if there 
was little or no evolution in the elliptical population to
redshifts of order unity. 
Thus a non-evolving population of elliptical galaxies is 
adopted here.

The mass profile of the deflectors has a greater impact on the 
frequency of multiply-imaged sources, as well as the resultant 
image configurations. 
Constant mass-to-light ratio de Vaucouleurs (1948) models of ellipticals 
can be matched to either the galaxy dynamics 
(Kormendy \& Djorgovski 1989; van der Marel 1991)
or lens statistics (Maoz \& Rix 1993; Kochanek 1996b), 
but not to both simultaneously. The mass-to-light ratios
required to reproduce the observed image separations are approximately
double those suggested by dynamical arguments.
This implies that ellipticals are dominated by dark matter 
halos, which might be expected to follow the
Navarro, Frenk \& White (1996, 1997) mass profile inferred from
$N$-body simulations of cold dark matter-dominated galaxy formation.
However, in the inner regions -- which determine the strong lensing
properties -- this profile
is only marginally steeper than the de Vaucouleurs (1948) model,
and is inconsistent with lensing observations for the same
reasons.
The inferred dark matter halos
can be modelled as isothermal spheres (\eg\ Binney \& Tremaine 1987), 
which 
are consistent with both dynamical 
considerations (\eg\ Kormendy \& Richstone 1995)
and lensing data (\eg\ Kochanek 1993, 1996b). 
There is, however, some uncertainty as to the correct mass normalisation
of this model (characterised by a velocity dispersion, $\sigma_\infty$), 
due to both its relationship with the observed
\los\ velocity dispersion,
$\sigma_{||}$, and the possibility of a finite core.
The lensing cross-section of an isothermal galaxy is proportional to
$\sigma_\infty^4$ (Turner \etal\ 1984), 
so even small variations in the normalisation
are important.

The surface brightness of ellipticals is flatter than 
an isothermal profile near the centre (and steeper at large radii, although 
this is less relevant), which results in higher observed 
velocity dispersions for a given mass distribution than constant
mass-to-light ratio models (\eg\ Binney \& Tremaine 1987). 
Gott (1977) used the conversion
$\sigma_\infty = (3/2)^{1/2} \sigma_{||}$ 
to account for the extended nature of the dark matter halo,
but more recent calculations imply that the correct scaling
is much closer to unity (\eg\ Kochanek 1993, 1994).
This result is supported by lens statistics (\eg\ Kochanek 1996b),
but the uncertainties 
are quite large.

If ellipticals do have finite
cores (within which the density is roughly constant), the 
the maximum deflection angle is reduced, making them less effective
lenses (\eg\ Hinshaw \& Krauss 1987; Blandford \& Kochanek 1987).
However, the mass normalisation is increased for a given 
observed $\sigma_{||}$,
as the central potential well of the galaxies are shallower;
this tends to increase their lensing effectiveness (Kochanek 1996a,b). 
Further, non-singular lenses
tend to produce more highly magnified images (\eg\ Blandford 
\& Kochanek 1987), resulting in an increased lensing probability
due to magnification bias (Turner 1980).
The qualitative arguments are quite clear, but the relative importance
of the various effects, and their overall impact on lens statistics,
are not.

In \sect{models} the normalisation and resultant
scalings of the mass distribution are derived, and the effects
these have on the optical depth and 
lensing probability of elliptical galaxies 
are discussed in 
\sect{probability}.
The conclusions reached on the effect of core radii on 
lens statistics are then summarised in \sect{conclusions}.

\section{Elliptical galaxies}
\label{section:models}

A simple model for the population of 
elliptical galaxies (\sect{population})
is adopted here,
in which
individual galaxies are assumed to be spherically symmetric objects, 
completely defined by their radial mass distribution 
(\sect{mass}), 
radial luminosity density (\sect{light}),
and dynamics.
Given that the line-of-sight velocity dispersion in the central regions
is usually used for the dynamical normalisation of ellipticals, 
this observable must be related to the model parameters for 
self-consistency (\sect{dynamics}).

\subsection{Population}
\label{section:population}

If elliptical galaxies follow a Schechter (1976)
luminosity function, and obey the Faber-Jackson (1976)
relationship,
their local co-moving number density is given by
\begin{equation}
\frac{{\rmn d} n_{\rmn g}}{{\rmn d} \sigma_{||}} =
\frac{\gamma n_*}{\sigma_*}
\left( \frac{\sigma_{||}}{\sigma_*}\right)^{\gamma(1 + \alpha) - 1}
\exp \left[ - \left( \frac{\sigma_{||}}
{\sigma_*}\right)^\gamma\right],
\label{equation:schechter}
\end{equation}
where 
$\sigma_{||}$ is the observed line-of-sight velocity dispersion,
$\alpha = - 1.07 \pm 0.05$ and 
$n_* = (0.0019 \pm 0.003)h^3$ Mpc$^{-3}$ (Efstathiou, Ellis \& Peterson 1988).
Here
$H_0 = 100h$ km s$^{-1}$ Mpc$^{-1}$
is Hubble's constant,
and 
$\sigma_* = 225 \pm 20$ km s$^{-1}$ and 
$\gamma = 3.7 \pm 1$ (de Vaucouleurs \& Olson 1982).
There is the possibility of systematic uncertainties 
in \eq{schechter} for low $\sigma_{||}$
(\eg\ Folkes \etal\ 1999), but the larger galaxies dominate the
strong lensing by ellipticals.
Further, any scatter in the Faber-Jackson (1976) relation
effectively increases $\sigma_*$ by an amount comparable to the scatter
in $\sigma_{||}$ (Kochanek 1994).

Under the assumption that the galaxy population is non-evolving
(See \sect{intro_core}.), 
the differential number of galaxies at redshift $z$ and 
and with velocity dispersion $\sigma_{||}$ is 
\begin{equation}
\label{equation:d2n_gal}
\frac{{\rmn d}^2 N_{\rmn g}}{{\rmn d}z \,{\rmn d}\sigma_{||}}
= \frac{{\rmn d}V_0}{{\rmn d}z} 
\frac{{\rmn d} n_{\rmn g}}{{\rmn d}\sigma_{||}},
\end{equation}
where ${\rmn d} V_0/{\rmn d} z$ is the co-moving volume element
at redshift $z$.
Its full cosmological dependence is rather complex (\eg\
Carroll, Press \& Turner 1992; Kayser, Helbig \& Schramm 1997)
and so only three simple, limiting cases are used here:
$\Omega_{\rm m_0} = 1$ and $\Omega_{\Lambda_0} = 0$
(the Einstein-de Sitter model);
$\Omega_{\rm m_0} = 0$ and $\Omega_{\Lambda_0} = 0$
(the empty Milne model);
and $\Omega_{\rm m_0} = 0$ and $\Omega_{\Lambda_0} = 1$
(a cosmological constant-dominated flat model).
In these models the volume element becomes
\begin{equation}
\label{equation:dv_0}
\frac{{\rmn d} V_0}{{\rmn d}z} =
\end{equation}
\[
\mbox{} \left\{
\begin{array}{lll}
4 \pi \left(\frac{c}{H_0}\right)^3
\frac{4 \left(z + 2 \sqrt{1 + z} + 2\right)}{(z + 1)^{5/2}},
& {\rmn if} & \Omega_{\rm m_0} = 1 \,\,{\rmn and}\,\, \Omega_{\Lambda_0}
= 0, \\ \\
4 \pi \left(\frac{c}{H_0}\right)^3
\frac{z^2 (z + 2)^2}{4 (z + 1)^3},
& {\rmn if} & \Omega_{\rm m_0} = 0 \,\,{\rmn and}\,\, \Omega_{\Lambda_0}
= 0, \\ \\
4 \pi \left(\frac{c}{H_0}\right)^3
z^2,
& {\rmn if} & \Omega_{\rm m_0} = 0 \,\,{\rmn and}\,\, \Omega_{\Lambda_0} = 1.
\end{array}
\right. 
\]

\subsection{Mass distribution}
\label{section:mass}

Despite its unbounded total mass, the 
non-singular isothermal sphere is consistent with 
the dynamics of elliptical galaxies (and their lensing properties).
The Hinshaw \& Krauss (1987) model has a 
mass density given by
\begin{equation}
\label{equation:rho_m}
\rho_{\rmn M} (r) = \frac{\sigma^2_\infty} {2 \pi G}
\frac{1}{(r^2 + r_{\rmn c}^2)},
\end{equation}
where $r_{\rmn c}$ is the core radius and 
$\sigma_\infty$ is the line-of-sight velocity dispersion away from 
the core for a constant mass-to-light ratio galaxy.
The integrated mass is given by 
\begin{equation}
\label{equation:m_r}
M(< r) = \frac{2 \sigma^2_\infty r_{\rmn c}}{G}
\left[\frac{r}{r_{\rmn c}} 
- \arctan \left(\frac{r}{r_{\rmn c}} \right) \right]
\end{equation}
and the projected surface density by
\begin{equation}
\label{equation:sig_m}
\Sigma_{\rmn M} (R) = \frac{\sigma^2_\infty}{2 G}
\frac{1}{\sqrt{R^2 + r_{\rmn c}^2}},
\end{equation}
which can also be integrated analytically to give
\begin{equation}
M(< R) = \frac{\sigma_\infty^2}{G} 
\left( \sqrt{R^2 + r_{\rmn c}^2} - r_{\rmn c}\right).
\label{equation:sig_inside}
\end{equation}

As discussed in \sect{light}, the luminosity
density of many ellipticals appears to be effectively singular,
but such observations cannot directly constrain the mass distribution.
Nonetheless, there are at least two strong arguments to suggest
that $r_{\rmn c}$ is small as well. Firstly, most
lensed quasars have even numbers of
images (\eg\ Keeton \& Kochanek 1996), which implies that
the galaxies' mass distributions are very nearly singular
(\eg\ Wallington \& Narayan 1993; Kassiola \& Kovner 1993).
Secondly,
dynamical modelling, combined with high-resolution
{\em Hubble Space Telescope} imaging
of nearby ellipticals, reveals that a number have
large black holes at their centres, and so are formally infinitely
dense there (\eg\ Kormendy \etal\ 1996, 1997).
If the core radius is non-zero, its scaling with 
$\sigma$ is important (Kochanek 1991; \sect{probability}), and is taken to be
\begin{equation}
\label{equation:c scaling}
r_{\rmn c} = r_{\rmn c*} \left( \frac{\sigma_{||}}{\sigma_*}
\right)^{u_{\rmn c}},
\end{equation}
where
$u_{\rmn c} = 4 \pm 1$ (\eg\ Fukugita \& Turner 1991).

\subsection{Light distribution}
\label{section:light}

Two models of the surface brightness of ellipticals are used here;
both are compatible with observations, and the difference between
the results of the two models is an indication of the uncertainty
of this calculation.

\begin{figure}
\includegraphics{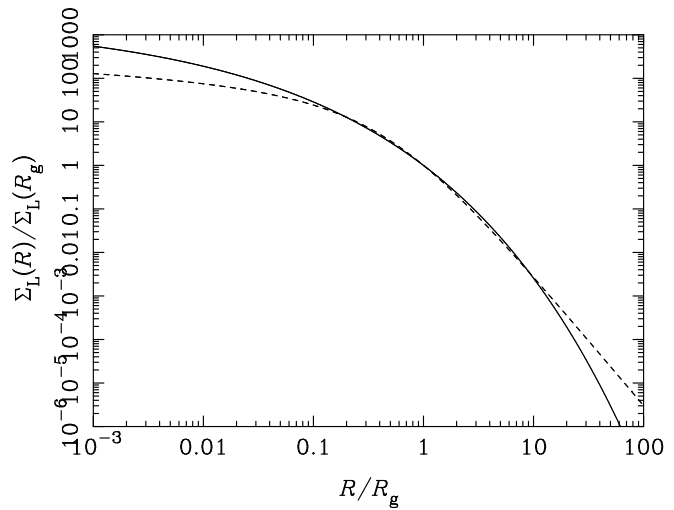}
\vspace{\singlefigureheight}
\caption{The projected luminosity density of galaxies described
by a de Vaucouleurs (1948) law (solid line)
and a Hernquist (1990) profile (dashed line), scaled by their values
at their effective or half-light radius, $R_{\rmn g}$.}
\label{figure:rho_lum_proj}
\end{figure}

The first is a de Vaucouleurs (1948) profile, given by
\begin{equation}
\label{equation:dev_sig_l}
\Sigma_{\rmn L} (R) = 
296.7 \frac{L}{\pi R_{\rmn g}^2}
\exp\left[{- 7.67 \left(\frac{R}{r_{\rmn g}}\right)^{1/4}}\right],
\end{equation}
where $R_{\rmn g}$ is the effective or half-light radius 
of the galaxy and $L$ is its luminosity. 
This is shown in \fig{rho_lum_proj} as the solid line.
The luminosity density is given by an 
Abel integral (Binney \& Tremaine 1987) as
\begin{equation}
\label{equation:abel}
\rho_{\rmn L} (r) = - \frac{1}{\pi} 
\int_r^\infty \frac{{\rmn d} \Sigma_{\rmn L}}
{{\rmn d} R} \frac{1}{\sqrt{R^2 - r^2}} \, {\rmn d}R.
\end{equation}
It must be computed numerically for most $r$,
but can be approximated by
$\rho_{\rmn L} (r) \simeq 1096.6\, L/(\pi^2 R_{\rmn g}^3)$
$(r / R_{\rmn g})^{-3/4}$ for $r \rightarrow 0$ (Young 1976).

The second model is based on a Hernquist (1990) profile,
which was developed as an analytical approximation to the
de Vaucouleurs (1948) profile,
but fits the data just as well in its own right.
With $R_{\rmn H} \simeq 0.55 R_{\rmn g}$,\footnote{The definition
$R_{\rmn H} \simeq 0.45 R_{\rmn g}$ is sometimes used (\eg\
Kochanek 1996b), but it is more relevant for the dynamics of 
constant mass-to-light models than for luminosity profiles alone.}
the surface brightness is 
\begin{equation}
\label{equation:her_sig_l}
\Sigma_{\rmn L} (R) = \frac{L}{2 \pi R_{\rmn H}^2}
\frac{\left[2 + (R / R_{\rmn H})^2\right]
\left[f_{\rmn H}(R / R_{\rmn H}) - 3\right]}
{\left[1 - (R / R_{\rmn H})^2\right]^2},
\end{equation}
where
\begin{equation}
f_{\rmn H}(x) = \left\{
\begin{array}{ccc}
\frac{{\rmn arccosh}(1 / x)}{\sqrt{1 - x^2}}, & {\rmn if} & x < 1, \\ \\
1, & {\rmn if} & x = 1, \\ \\
\frac{\arccos(1 / x)}{\sqrt{x^2 - 1}}, & {\rmn if} & x > 1.
\end{array}
\right.
\end{equation}
This is shown as the dashed line in \fig{rho_lum_proj}, 
and is qualitatively similar to the de Vaucouleurs (1948)
profile for $0.1 R_{\rmn g} \la R \la 10 R_{\rmn g}$.
The fraction of the flux in the discrepant regions is 
only a few per cent.
The resultant luminosity density is given by 
\begin{equation}
\label{equation:her_rho_l}
\rho_{\rmn L} (r) = \frac{L}{2 \pi R_{\rmn H}^3}\frac{1}
{r / R_{\rmn H} (1 + r / R_{\rmn H})^3}.
\end{equation}

The spatial scale of both distributions is determined by the effective
radius, which, like the core radius of the mass distribution, 
is assumed to increase with the 
velocity dispersion of the galaxy as
\begin{equation}
\label{equation:g scaling}
R_{\rmn g} = R_{\rmn g*} 
\left(\frac{\sigma_{||}}{\sigma_*}\right)^{u_{\rmn g}},
\end{equation}
where $R_{\rmn g*} = (4 \pm 1) h$ kpc and $u_{\rmn g} = 4 \pm 1$
(Kormendy \& Djorgovski 1989).

\subsection{Dynamical normalisation}
\label{section:dynamics}

The population of galaxies is given in terms of $\sigma_{||}$ 
in \sect{population}, but the mass distribution,
and hence lensing properties of individual galaxies are 
determined by $\sigma_\infty$ (\sect{mass}).
For a given $\sigma_\infty$, the central dispersion decreases with
increasing core radius, but also depends on both the stellar 
dynamics within the galaxy and the luminosity profile. 
Radial orbits result in higher dispersions, as the stars fall
through the core of the galaxy, and more extended luminosity
profiles also result in faster central stellar motions.
The galaxy model is specified by $\rho_{\rmn M}(r)$
and $\rho_{\rmn L}(r)$, given in \sects{mass}
and \ref{section:light}, respectively,
and the (assumed constant) velocity anisotropy, $\beta_\sigma$. 
This is defined as $\beta_\sigma 
= 1 - \sigma_\theta^2 / \sigma_r^2$, 
where $\sigma_\theta$, $\sigma_\phi$ (which are equal in a 
non-rotating system) and $\sigma_r$ are the angular
and radial components, respectively,
of the velocity dispersion tensor of
the luminous matter. Both theoretical and observational
results suggest that $0.0 \la \beta_\sigma \la 0.5$
for ellipticals
(Binney \& Tremaine 1987; van der Marel 1991; Kochanek 1994), but
a broader range of values is explored here.

Under the above assumptions, the Jeans equation reduces to
(\eg\ Binney \& Tremaine 1987)
\begin{equation}
\sigma_r^2(r) \left[
\frac{2}{\sigma_r(r)} \frac{{\rmn d} \sigma_r}{{\rmn d}r}
+ \frac{1}{\rho_{\rmn L}(r)} \frac{{\rmn d} \rho_{\rmn L}}{{\rmn d}r}
+ \frac{2 \beta_\sigma}{r}
\right]
= - \frac{G M(< r)}{r^2},
\end{equation}
which can be integrated to give
\begin{equation}
\sigma_r (r) = \left[ \frac{r^{- 2 \beta_\sigma}}
{\rho_{\rmn L} (r)} 
\int_r^\infty G M(< r^\prime) \rho_{\rmn L}(r^\prime) 
{r^\prime}^{2(\beta_\sigma - 1)} 
\, {\rmn d} r^\prime
\right]^{1/2}.
\end{equation}
This can be found analytically in the regions where $\rho_{\rmn M}$
is purely isothermal and $\rho_{\rmn L}$ can be approximated 
by a global power law, given by $\rho_{\rmn L}(r) \propto r^{-\xi}$
[and hence $\Sigma_{\rmn L} \propto R^{- (\xi - 1)}$,
provided that $\xi > 1$]. 
The Hernquist (1990) model approaches $\xi = 1$ for 
small $r$ and $\xi = 4$ for large $r$; the de Vaucouleurs (1948)
law has $\xi = 3/4$ for small radii.
The radial dispersion in these regimes is then
\begin{equation}
\label{equation:sigma_r_power}
\sigma_r (r) = \sigma_\infty \left(\frac{2}{\xi - 2 \beta_\sigma}
\right)^{1/2}.
\end{equation}
In the outer
regions the core radius has no effect, but for $r \la r_{\rmn c}$
the lowered central density is more important than both
$\rho_{\rmn L}$ and the velocity anisotropy (assuming
$\beta_\sigma \la 0.5$). If $\beta_\sigma \simeq 1$, the orbits
are predominantly radial, and luminous material `plunges'
through the central regions of the galaxy, greatly increasing $\sigma_r$.

\begin{figure*}
\includegraphics{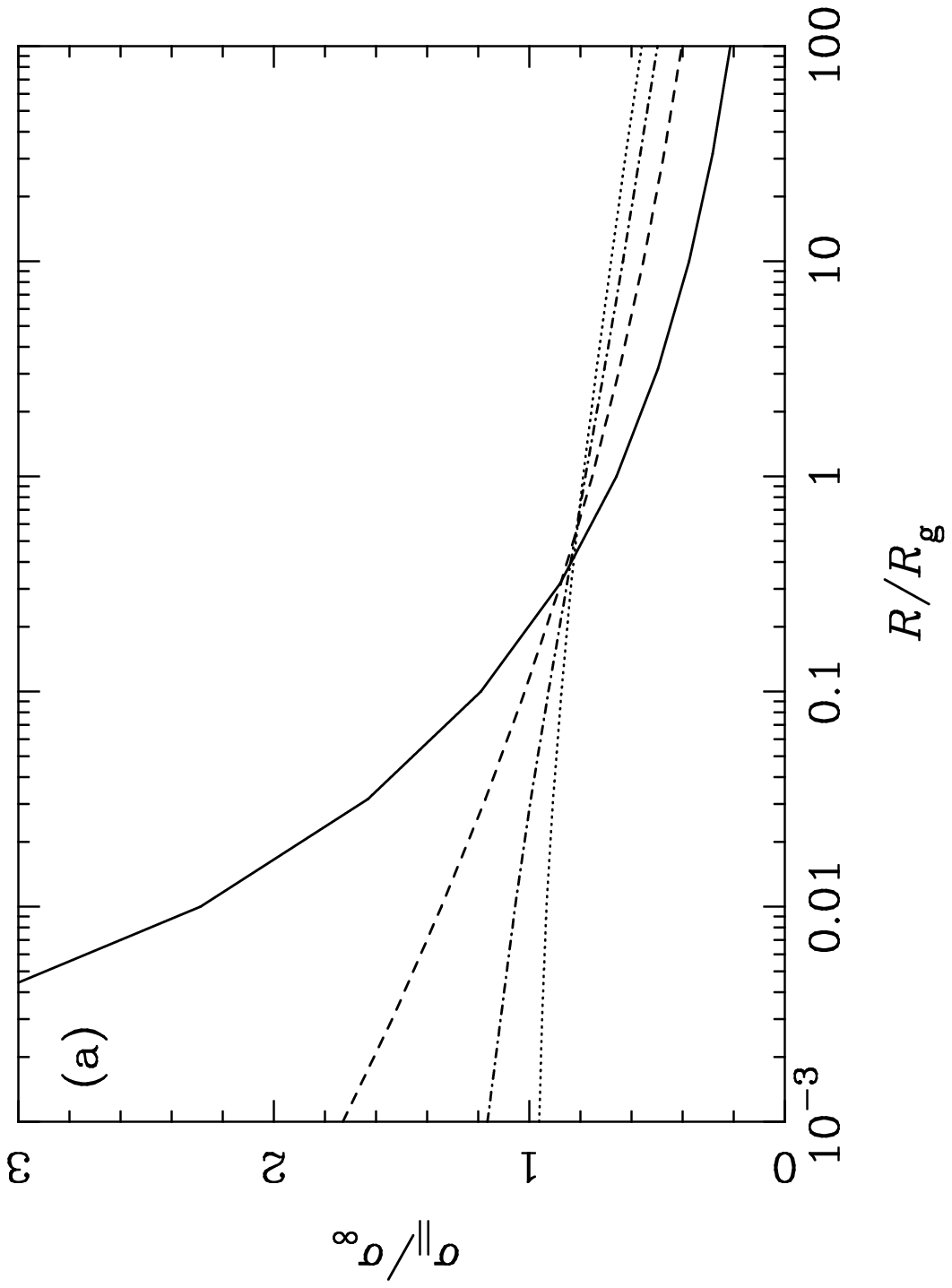}
\includegraphics{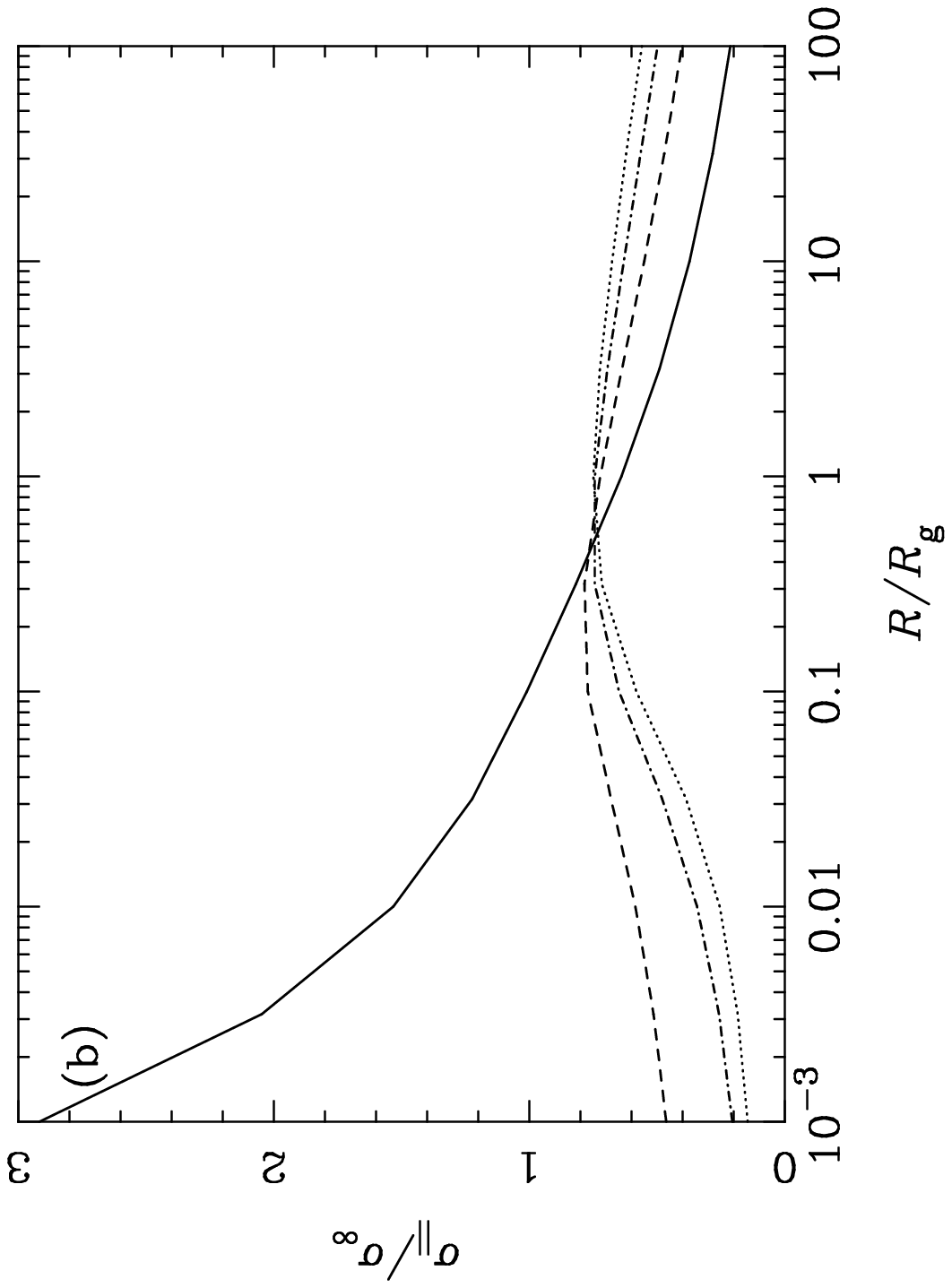}
\includegraphics{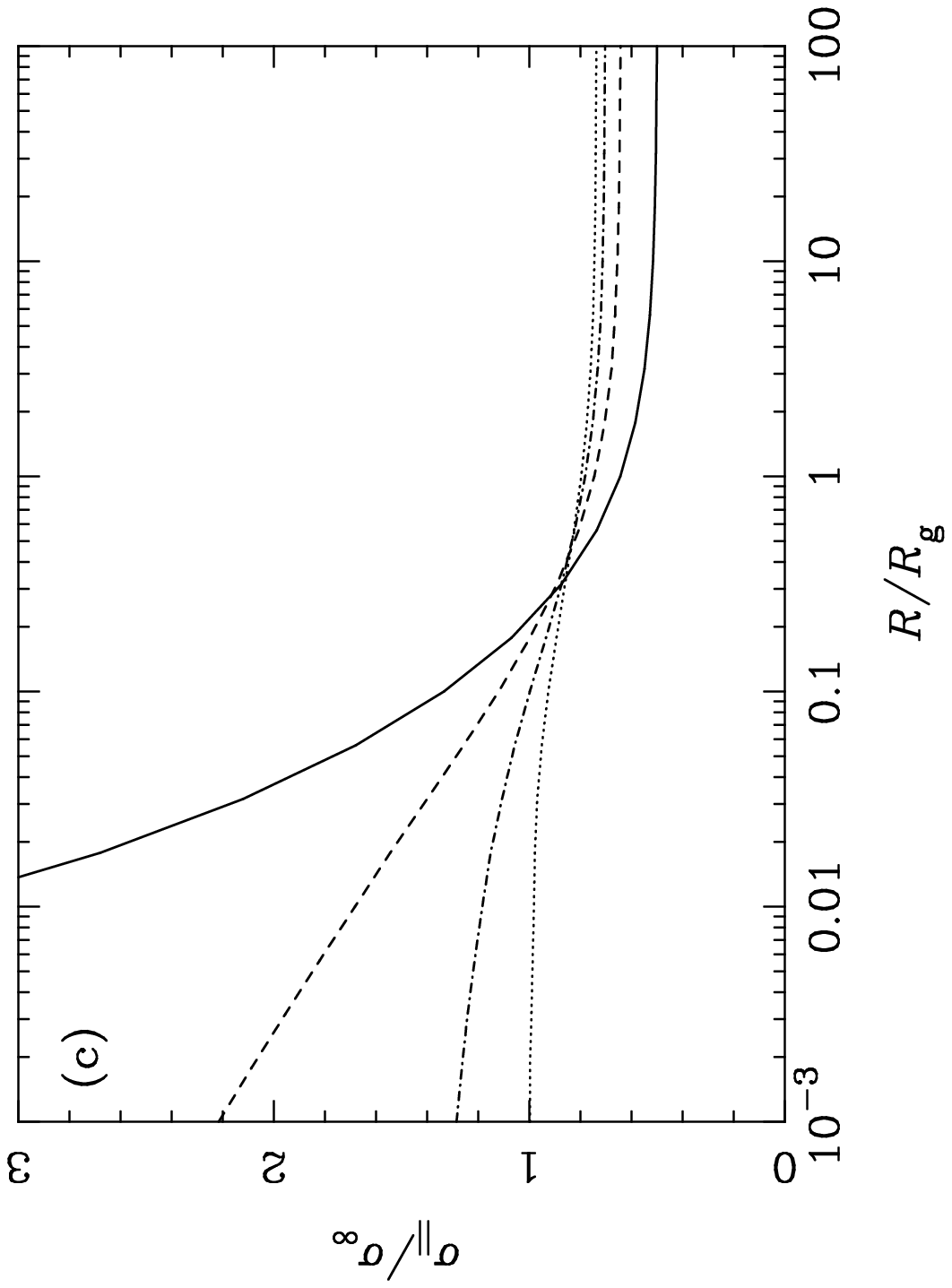}
\includegraphics{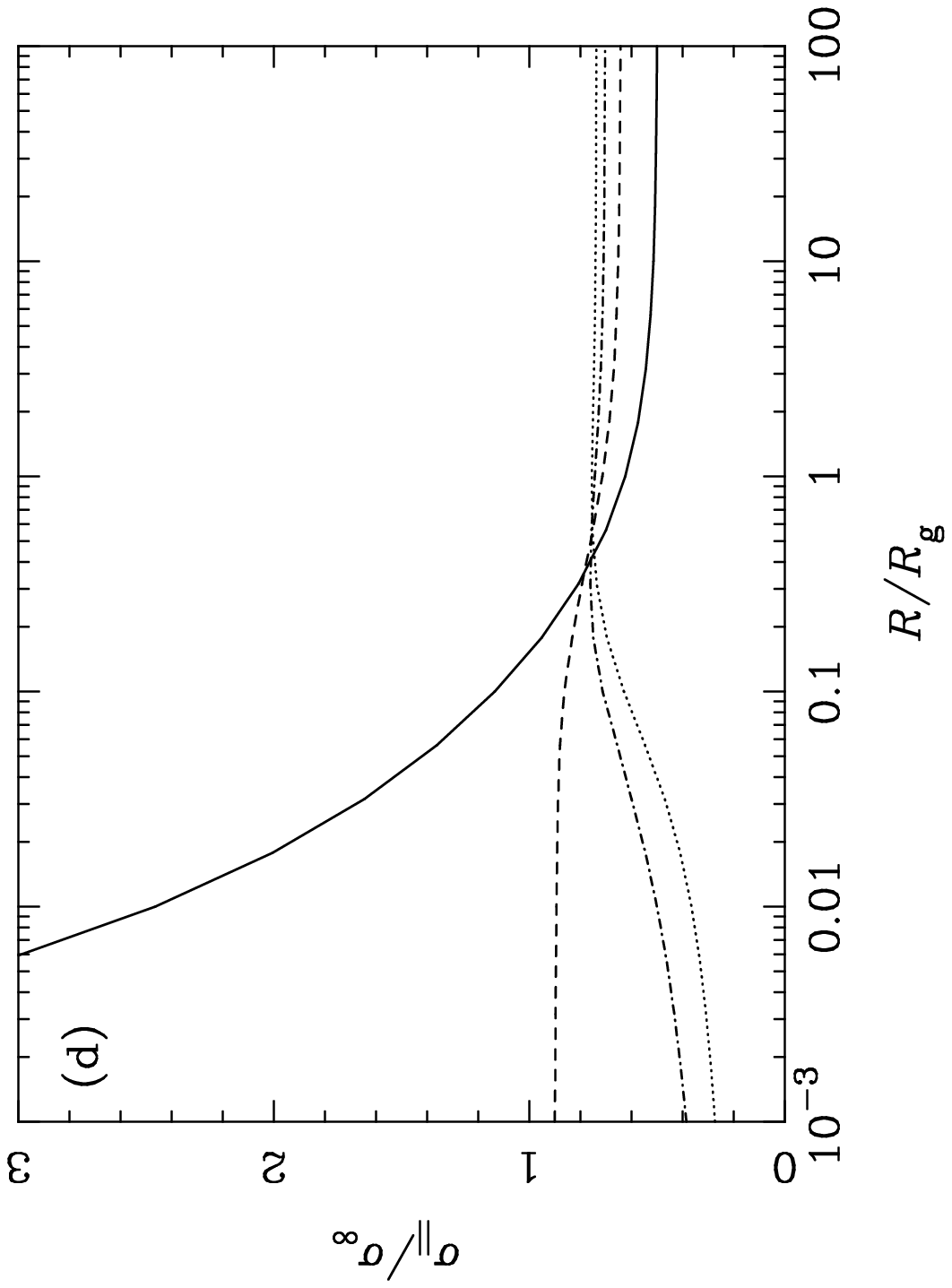}
\vspace{\doublefigureheight}
\caption{The line-of-sight velocity dispersion, $\sigma_{||}$,
of the luminous matter in an elliptical galaxy with
luminosity densities described by a de Vaucouleurs (1948) law
in (a) and (b) and a Hernquist (1990) model in (c) and (d).
In (a) and (c) the mass distribution is a singular
isothermal sphere; in (b) and (d) it has a core radius
of $0.1 R_{\rmn g}$, where $R_{\rmn g}$ is the effective radius 
of the galaxy.
The four lines in each panel represent different values of
the velocity anisotropy: $\beta_\sigma = 1.0$ (solid lines);
$\beta_\sigma = 0.5$ (dashed lines); 
$\beta_\sigma = 0.0$ (dot-dashed lines);
and $\beta_\sigma = -0.5$ (dotted lines). The regions where $\sigma_{||}$ 
is flat
are those where the luminosity density can be approximated as
a global power law.}
\label{figure:sigma_los}
\end{figure*}

The line-of-sight velocity dispersion at a given position
on the galaxy is given by a projection integral (\eg\ Binney \& 
Tremaine 1987) as
\[
\sigma_{||}(R) = 
\left[ \frac{2}{\Sigma_{\rmn L} (R)}
\int_R^\infty \frac{(1 - \beta_\sigma 
R^2 / r^2) r}{\sqrt{r^2 - R^2}} \,
\rho_{\rmn L} (r) \sigma_r^2(r) \, {\rmn d} r \right]^{1/2}.
\]
\begin{equation}
\label{equation:sigma_los}
\end{equation}
The power law approximations discussed above are also valid here;
insertion of \eq{sigma_r_power} into the 
integral yields
\begin{equation}
\sigma_{||}(R) = \sigma_\infty \left[\frac{2}{\xi - 2 \beta_\sigma}\,
\frac{\xi - (\xi - 1) \beta_\sigma}{\xi}
\right]^{1/2} .
\label{equation:sigma_los_power}
\end{equation}
In the special case of $\xi = 3$, the standard result (Gott 1977), 
valid for all $\beta_\sigma$, that
$\sigma_{||}(R) = (2/3)^{1/2} \sigma_\infty$,
is recovered.
It is also clear that the line-of-sight dispersion is lower
in regions with a steeper luminosity density.
\Fig{sigma_los} shows $\sigma_{||}/\sigma_{\infty}$ as a
function of $R$ for the 
de Vaucouleurs (1948) and Hernquist (1990) models. 
For both singular models and those with $\beta_\sigma \simeq 1$, 
the central dispersion results is much higher than the
dispersions at large 
radii. Conversely, the central dispersion decreases with core radius, 
as the galaxy's gravitational well is shallower, and bound orbits 
must be slower. Again the core
radius is more important than the anisotropy of luminosity 
profile, provided only that $\beta_\sigma \la 0.5$.

\begin{figure*}
\includegraphics{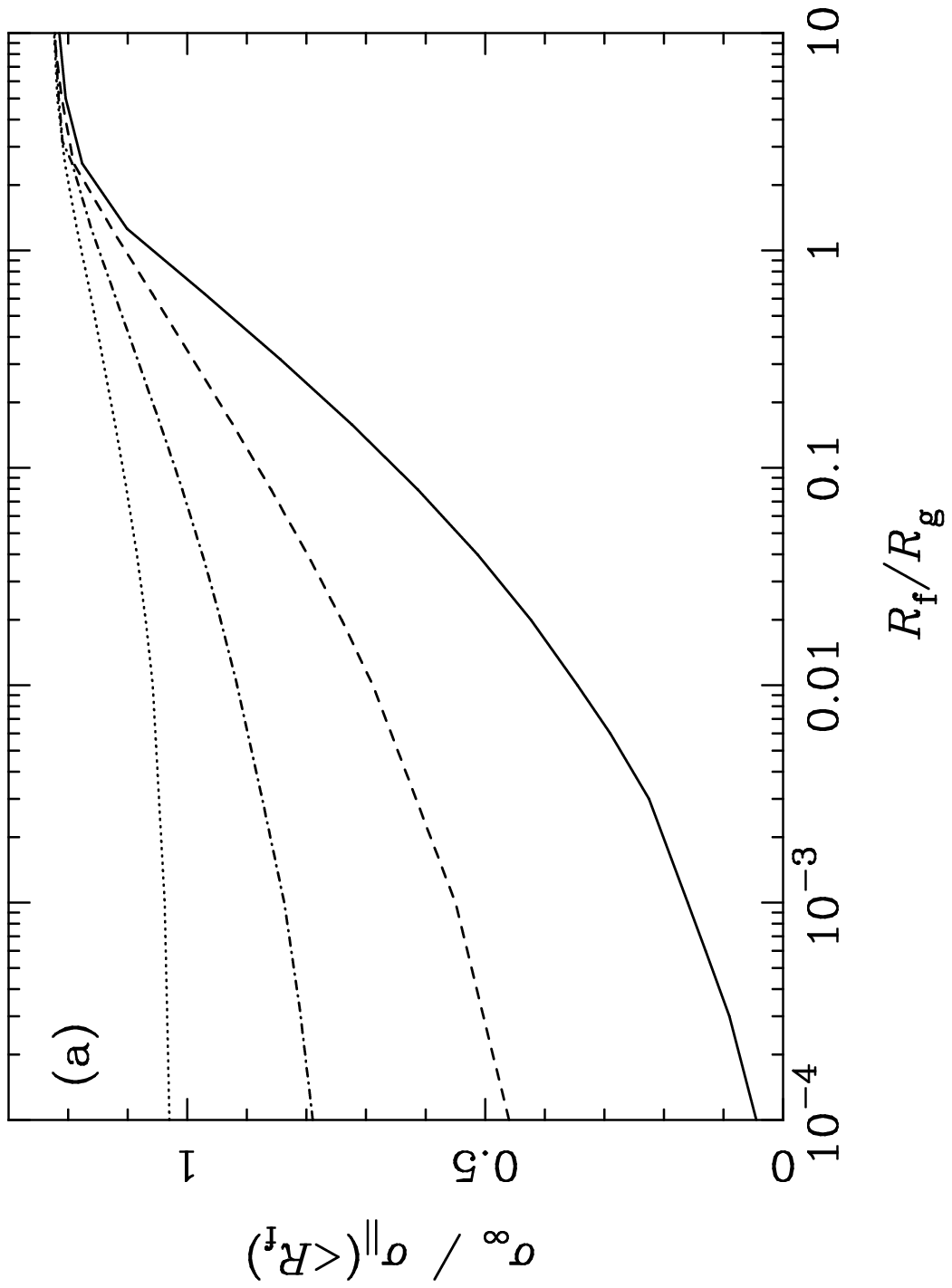}
\includegraphics{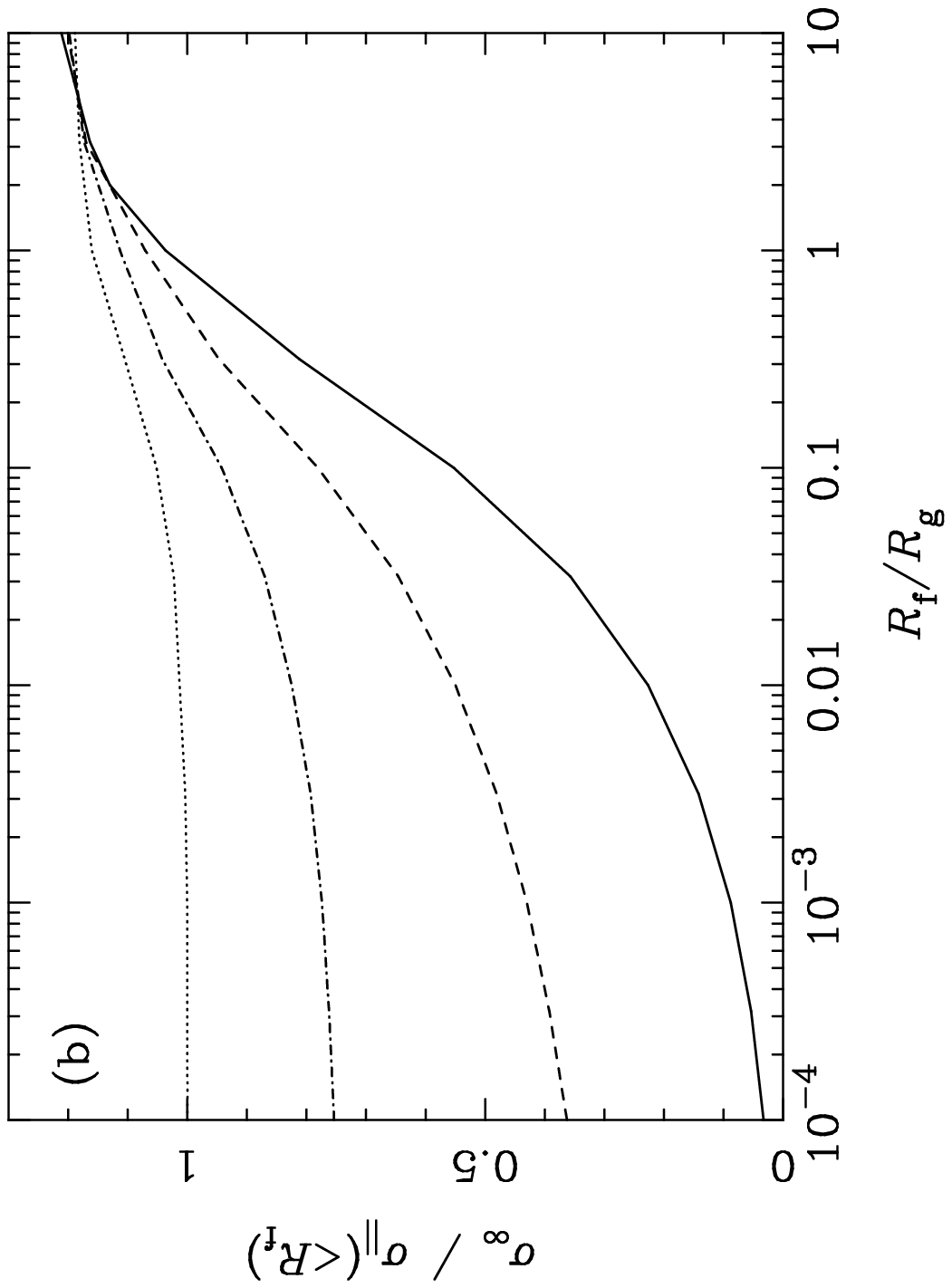}
\vspace{\singlefigureheight}
\caption{The dynamical normalisation, $\sigma_\infty$, as a
function of the size of the integration aperture, $R_{\rmn f}$,
for a singular galaxy with effective radius $R_{\rmn g}$
and observed line-of-sight velocity dispersion $\sigma_{||}$.
The luminosity density is described by a de Vaucouleurs (1948) law
in (a) and a Hernquist (1990) model in (b).
Different values of
the velocity anisotropy are denoted by the
different line-styles: $\beta_\sigma = 1.0$ (solid lines);
$\beta_\sigma = 0.5$ (dashed lines); $\beta_\sigma = 0.0$ (dot-dashed lines);
and $\beta_\sigma = -0.5$ (dotted lines).}
\label{figure:sigma_los_inside_fib}
\end{figure*}

\begin{figure*}
\includegraphics{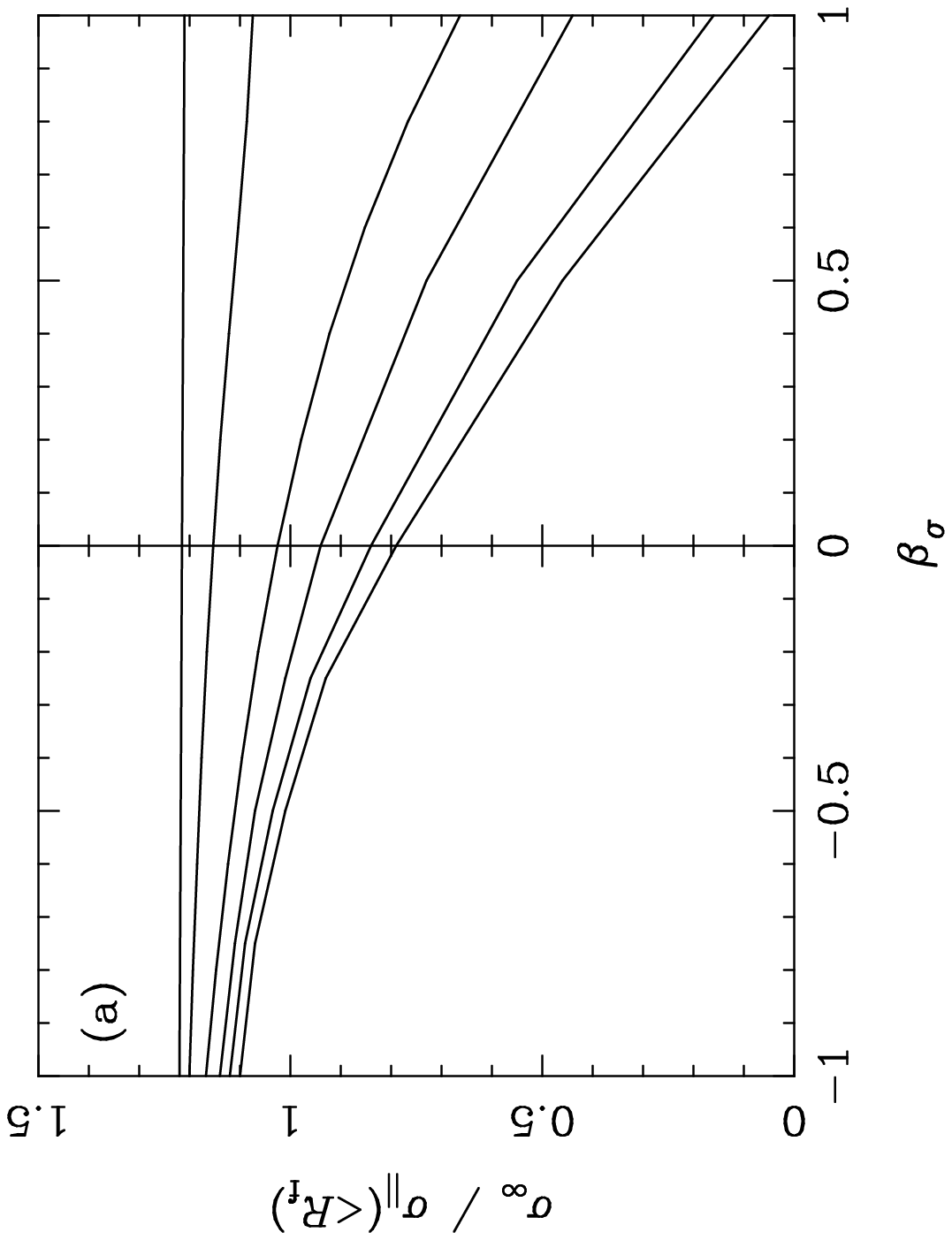}
\includegraphics{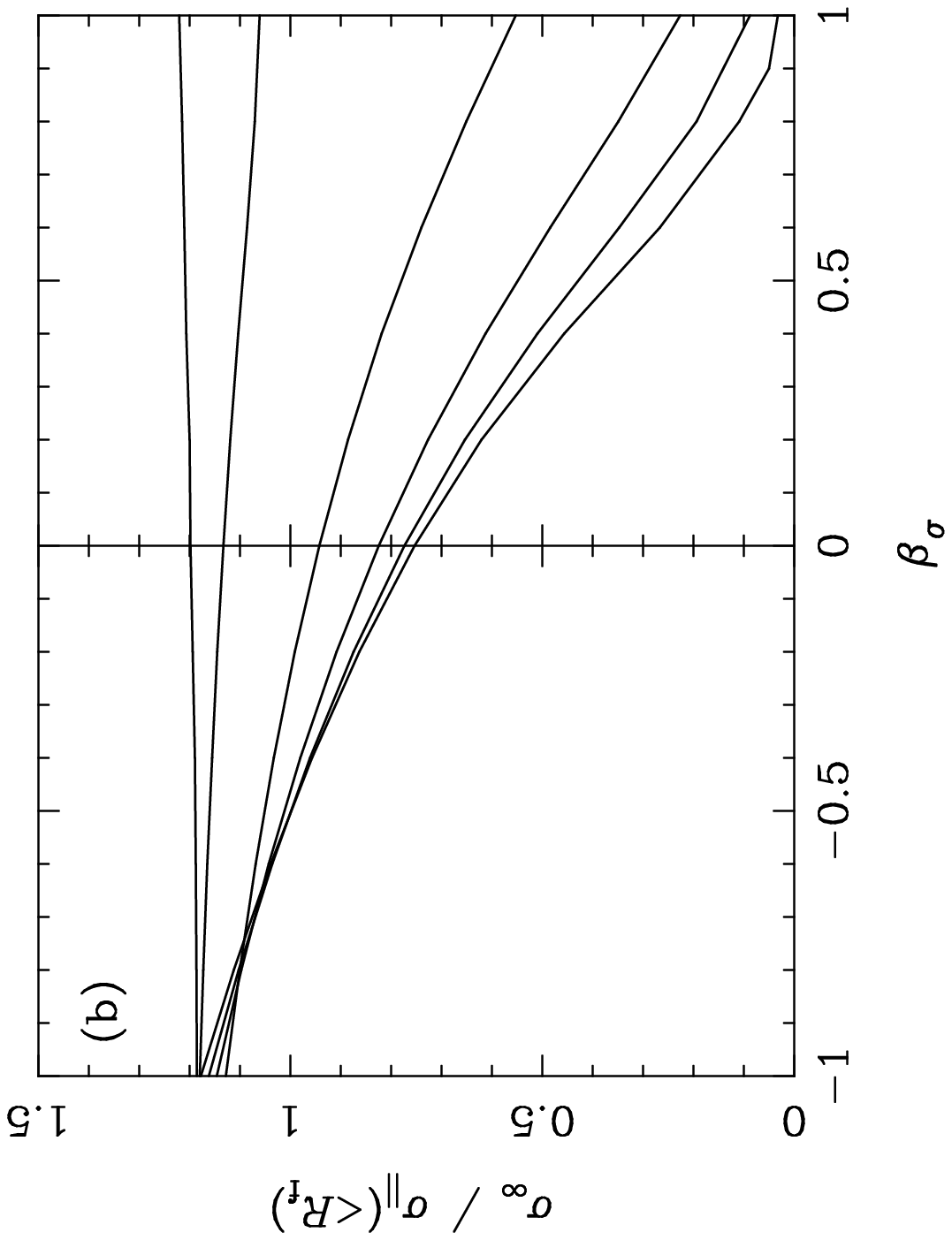}
\vspace{\singlefigureheight}
\caption{The dynamical normalisation, $\sigma_\infty$, as a
function of the velocity anisotropy, $\beta_\sigma$,
for a singular galaxy with effective radius $R_{\rmn g}$
and observed line-of-sight velocity dispersion $\sigma_{||}$.
The luminosity density is described by a de Vaucouleurs (1948) law
in (a) and a Hernquist (1990) model in (b).
The lines represent different aperture sizes -- from the bottom
to the top, $R_{\rmn f}/R_{\rmn g}$ takes on the values:
0.0001, 0.001, 0.01, 0.1, 1.0 and 10.0.}
\label{figure:sigma_los_inside_beta}
\end{figure*}

\begin{figure*}
\includegraphics{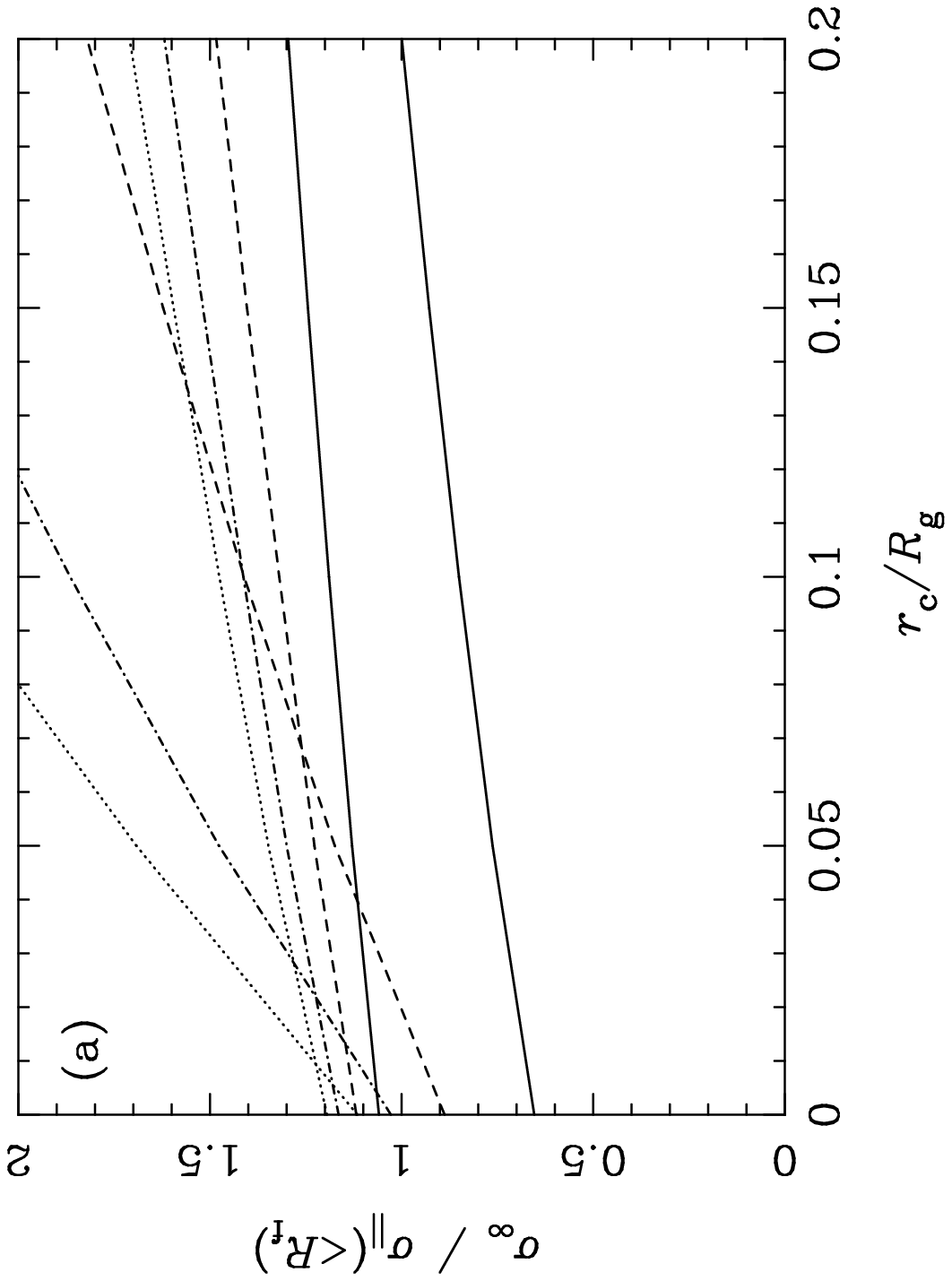}
\includegraphics{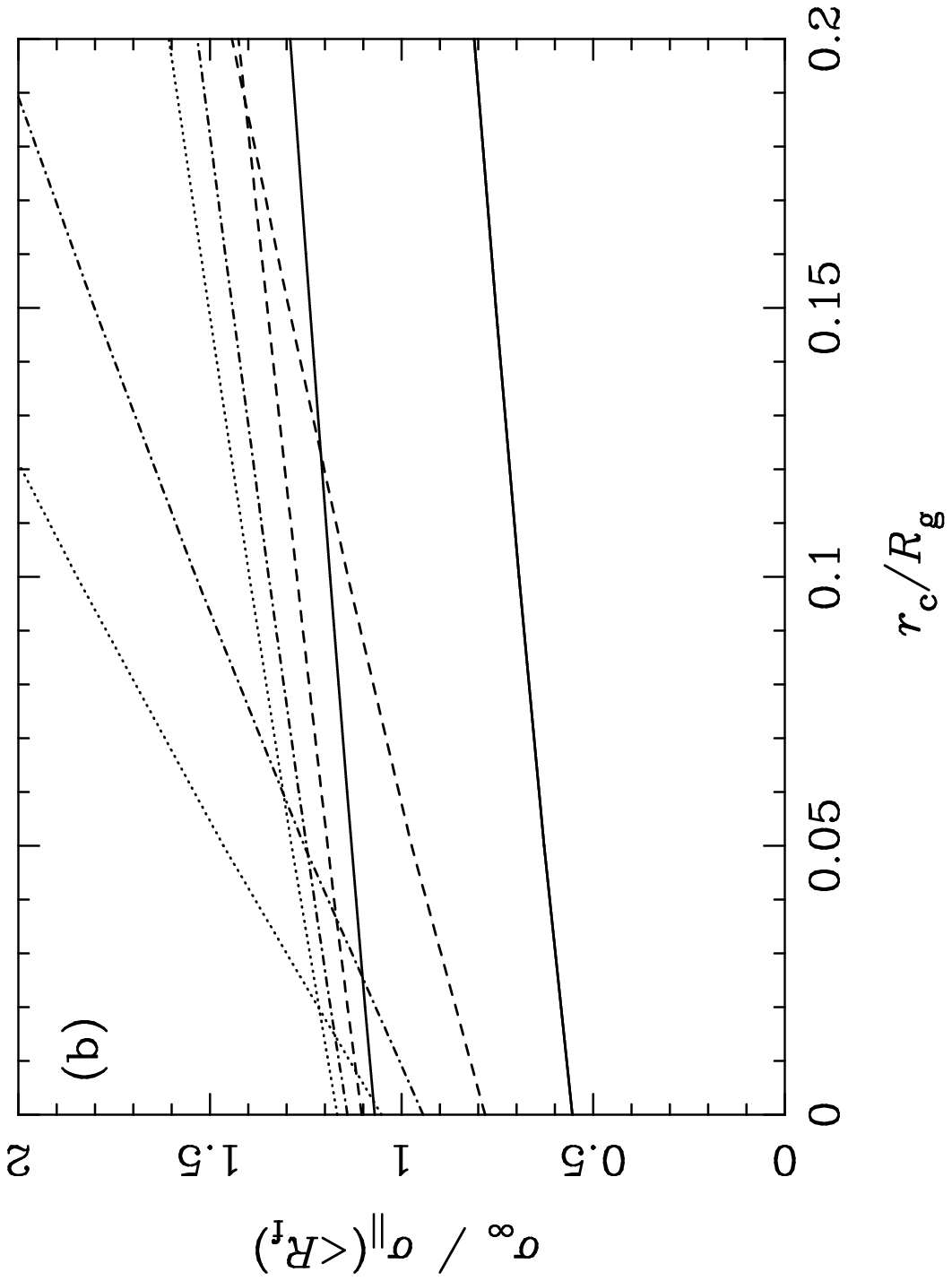}
\vspace{\singlefigureheight}
\caption{The dynamical normalisation, $\sigma_\infty$, as a function
of core radius, $r_{\rmn c}$, for a galaxy with effective
radius $R_{\rmn g}$ and observed line-of-sight velocity 
dispersion $\sigma_{||}$.
The luminosity density is described by a de Vaucouleurs (1948) law
in (a) and a Hernquist (1990) model in (b).
Different values of
the velocity anisotropy are denoted by the
different line-styles: $\beta_\sigma = 1.0$ (solid lines);
$\beta_\sigma = 0.5$ (dashed line); 
$\beta_\sigma = 0.0$ (dot-dashed lines);
and $\beta_\sigma = -0.5$ (dotted lines).
The two lines for each dynamical model are
for different values of the aperture size:
$R_{\rmn f} = 0.1 R_{\rmn g}$; and $R_{\rmn f} = R_{\rmn g}$.
The larger projected fibre radius results in
less variation between dynamical models and
higher values of
$\sigma_\infty$ for $r_{\rmn c} = 0$, but lower values of
$\sigma_\infty$ if the
core radius is large.}
\label{figure:sigma_los_inside}
\end{figure*}

A real velocity dispersion is measured from a spectrum taken over 
a finite region of the galaxy, in the form of either a linear slit 
or, more commonly now, an optical fibre.
Assuming a circular aperture of projected radius $R_{\rmn f}$
centred on the galaxy,
the observed line-of-sight velocity dispersion is 
\begin{equation}
\sigma_{||} (< R_{\rmn f}) = 
\frac{\int_0^{R_{\rmn f}} 2 \pi R \, \Sigma_{\rmn L} (R)
\sigma_{||}(R) \, {\rmn d}R }
{
\int_0^{R_{\rmn f}} 2 \pi R \, \Sigma_{\rmn L} (R)\, {\rmn d}R}.
\end{equation}
Atmospheric seeing would have the effect of smoothing the
function $\sigma_{||}(R)$, but its importance is minimal as 
this tends to be a slowly varying function for $R \ga R_{\rmn g}$.
For a fixed angular size, $\theta_{\rmn f}$,
the aperture will vary from galaxy to galaxy as $R_{\rmn f} = d_{\rmn A}(0, z)
\theta_{\rmn f} \simeq c z \theta_{\rmn f} / H_0$, 
where $z \ll 1$ is the galaxy's redshift, 
and $d_{\rmn A}$ its the angular diameter distance.
Only nearby galaxies are practical targets for dynamical
studies -- for instance the samples of van der Marel (1991)
and Lauer \etal\ (1995)
contain galaxies with $0.002 \la z \la 0.02$
and $0.0006 \la z \la 0.06$, respectively. 
Hence an aperture a few arcsec in diameter implies 
$0.01$ kpc $\la R_{\rmn f} \la 1$ kpc, as compared to 
core radii of $\sim 0.1$ kpc 
and
effective radii of $\simeq 4$ kpc.
\Figs{sigma_los_inside_fib} and \ref{figure:sigma_los_inside_beta}
show the normalisation for singular models as a function of
$R_{\rmn f}$ and $\beta_\sigma$, respectively.
The normalisation increases with $R_{\rmn f}$ as the
observed dispersion is less dependent on the extreme orbital speeds near
the core. Conversely, the normalisation decreases with $\beta_\sigma$,
as the maximum orbital speeds are higher for radial orbits.
\Fig{sigma_los_inside}
shows $\sigma_\infty / \sigma_{||}(< R_{\rmn f})$
as a function of core radius.
The general trend is that the observed velocity dispersion 
decreases (and so the dynamical normalisation increases)
with core radius and also with $R_{\rmn f}$. It is also
clear that the quantitative behaviour varies comparably with
the dynamical model and luminosity density assumed. 
For smaller apertures this dependence is more extreme,
as only the core region is directly observed.

Calculation of $\sigma_{||}(< R_{\rmn f})$ is computationally expensive,
especially for the de Vaucouleurs (1948) model; 
fortunately its dependence on core radius is very nearly linear. 
For the subsequent lens calculations, the normalisation used 
is 
\begin{equation}
\sigma_\infty = \sigma_{||}(< R_{\rmn f}) \,
\left( A + B
\frac{r_{\rmn c}}{R_{\rmn g}} \right),
\label{equation:sig_inf}
\end{equation}
where $A$ and $B$ independent of core radius.
Kochanek (1996b) used $A = 1$ and $B = 2$
for galaxies with $R_{\rmn f} \simeq R_{\rmn g}$,
but 
this underestimates the normalisation by up to $\sim 15$ per cent
(See \fig{sigma_los_inside}.),
and hence underestimates the lensing probability by up to $\sim 80$ per cent.
As defined above, $A$ is determined purely by the singular models,
for which the normalisation in plotted in 
\figs{sigma_los_inside_fib} and \ref{figure:sigma_los_inside_beta}.
Also, $A$ should match the 
power-law approximation given by \eq{sigma_los_power} 
for very small apertures,
and $\sigma_\infty / \sigma_{||} (< R_{\rmn f})
\rightarrow (3/2)^{1/2} \simeq 1.225$ for $R_{\rmn f} \rightarrow \infty$,
as shown by Kochanek (1993).
For intermediate apertures, 
the variation between the results for the two luminosity
profiles can be used as a guide to the uncertainty in the above results.
Given this discrepancy, analytical expressions for $A$ and $B$ were 
developed for both the de Vaucouleurs (1948) and Hernquist (1990)
profiles, and 
these were used in the lensing calculations presented in \sect{probability}.
The analytical forms are given in Mortlock (1999), and, 
for all reasonable galactic
models, agree with the numerical results to within a
few per cent; the 
uncertainty in the normalisation caused by the 
ambiguity in the choice of luminosity profile is considerably
larger at around 10 per cent.

Having found a relationship between the isothermal mass 
normalisation and observables for a given galaxy, there
still remains the question of how best to make the conversion
from the observed Faber-Jackson (1976) relation.
A `natural' choice is to consider the aperture to
be a specified fraction of the effective radius of each galaxy in
the sample although this is unrealistic for large surveys undertaken 
using fibres of a fixed radius.
(However, if the data is available, it is 
preferable to measure the velocity dispersion
out to $\sim R_{\rmn g}$, as the results are more robust.)
The other possibility is to treat measured dispersions 
as being averaged over a given physical scale, 
independent of $\sigma_{||}$, but this breaks down 
for small galaxies as only the central regions are likely to
be registered due to surface brightness considerations.
One way to circumvent these ambiguities is to fit the mass model 
directly from dynamical measurements on a 
galaxy-by-galaxy basis (\eg\
van der Marel 1991; Kockanek 1994). This is not
only more time-consuming, but also requires spatially resolved
surface brightness and velocity dispersion measurements.

The `default' model ($\beta_\sigma = 0$; $R_{\rmn f} \simeq 
R_{\rmn g}$; $r_{\rmn c} = 0$) has $\sigma_\infty / \sigma_{||}
\simeq 1.1$, leading to a 50 per cent increase in the lensing
probability,
due to its $\sigma_\infty^4$ dependence
(\eg\ Turner \etal\ 1984; Kochanek 1994). 
For most of the other plausible sets of parameter values the 
differences are even greater. 
Similar increases in the lensing probability can also result from 
the spread in the Faber-Jackson (1976) relation -- 
a dispersion of $\Delta \sigma_\infty / \sigma_\infty \simeq 0.2$ 
increases the lensing optical depth (\sect{tau}) by
a factor of $\sim 1 + 6 (\Delta \sigma_\infty / \sigma_\infty)^2 \simeq 1.25$
(Kochanek 1994).
If a number of other common simplifications used in lensing 
calculations are correct (\eg\ that spiral galaxies are 
unimportant; that obscuration by dust is minimal; etc.), 
the dynamical normalisation is probably the greatest 
uncertainty in lens statistics at present. 
Further, whereas uncertainty in $n_*$ can be greatly decreased 
by simple counting procedures, accurate normalisation (or,
equivalently, measurement of ${\rmn d}n_{\rmn g}/{\rmn d}\sigma_\infty$)
requires much more detailed data.

\section{Lensing probability}
\label{section:probability}

From the introduction of finite cores to lens galaxies by
Hinshaw \& Krauss (1987) 
it was generally believed that the
fraction of quasars which are multiply-imaged drops sharply
with increasing core radius. However as shown by 
Kochanek (1996a,b) the requirements of self-consistent dynamical 
normalisation (\sect{dynamics}) imply that the lensing
probability declines only slowly, or possibly even increases with core 
radius. Once the lens equation has been solved (\sect{lens eq}),
these possibilities are quantified in terms of both
the optical depth (\sect{tau})
and a simple but self-consistent calculation of lensing probability
of a distant quasar (\sect{prob}).

\subsection{The lens equation}
\label{section:lens eq}

The lens equation relates the angular position (relative to the
optical axis joining the observer and the centre of the lens)
of the source, $\beta$, to those of
its image(s), $\theta$. These quantities can only be related via
$d_{\rmn{A}} (0, z_{\rmn{d}})$, $d_{\rmn{A}} (0, z_{\rmn{s}})$ and
$d_{\rmn{A}} (z_{\rmn{d}}, z_{\rmn{s}})$, the angular
diameter distances
from observer to deflector,
observer to source, and deflector to source, respectively.
In the filled-beam approximation (Dyer \& Roeder 1972, 1973),
the angular diameter distances in the cosmological models 
described in \sect{population} are given by
\begin{equation}
\label{equation:d_a_core}
d_{\rmn A}(z_1, z_2) = 
\end{equation}
\[
\mbox{}
\left\{
\!\!\!
\begin{array}{lll}
\frac{c}{H_0} \frac{2}{z_2 + 1} 
\left(\frac{1}{\sqrt{z_1 + 1}} - \frac{1}{\sqrt{z_2 + 1}} \right), \!\!\!\!
& {\rmn if} & \Omega_{\rm m_0} = 1 \,\,{\rmn and}\,\, \Omega_{\Lambda_0}
= 0, \\ \\
\frac{c}{H_0} \frac{z_2 (z_2 + 2) - z_1 (z_1 + 2)}
{2 (z_1 + 1) (z_2 + 1)^2}, \!\!\!\!
& {\rmn if} & \Omega_{\rm m_0} = 0 \,\,{\rmn and}\,\, \Omega_{\Lambda_0}
= 0, \\ \\
\frac{c}{H_0} \frac{z_2 - z_1}{z_2 + 1}, \!\!\!\!
& {\rmn if} & \Omega_{\rm m_0} = 0 \,\,{\rmn and}\,\, \Omega_{\Lambda_0} = 1 .
\end{array}
\right. 
\]
For a thin lens with a cumulative surface mass distribution 
$M (< R)$, standard techniques 
(\eg\ Schneider, Ehlers \& Falco 1992)
give the lens equation as
\begin{equation}
\beta = \theta - \frac{1}{\pi \theta} 
\frac{M [< d_{\rmn A}(0, z_{\rmn d}) \theta]}
{d_{\rmn A}^2(0, z_{\rmn d}) \, \Sigma_{\rmn crit} (z_{\rmn d}, z_{\rmn s})},
\end{equation}
where 
\begin{equation}
\label{equation:sigma_crit}
\Sigma_{\rmn crit} (z_{\rmn d}, z_{\rmn s})
= \frac{c^2}{4 \pi G} \frac{d_{\rmn A}(0, z_{\rmn s})}
{d_{\rmn A}(0, z_{\rmn d}) \, d_{\rmn A}(z_{\rmn d}, z_{\rmn s})}
\end{equation}
is the surface density required for a rotationally-symmetric
lens to be capable of forming multiple images
(\eg\ 
Subramanian \& Cowling 1986; Schneider \etal\ 1992).

Using the mass distribution
given in \eq{sig_inside}, 
the lens equation becomes 
\begin{equation}
\label{equation:dimensionless lens equation}
\beta = \theta - \theta_{\rmn{E}}
\frac{\sqrt{1 + \theta^2 / \theta_{\rmn{c}}^2} - 1}
{\theta / \theta_{\rmn{c}}}.
\end{equation}
where $\theta_{\rmn{c}} = r_{\rmn{c}} /
d_{\rmn{A}} (0, z_{\rmn{d}})$ and
\begin{equation}
\label{equation:einstein angle}
\theta_{\rmn{E}} = 4 \pi
\left(\frac{\sigma_\infty}{c}\right)^2
\frac{d_{\rmn{A}} (z_{\rmn{d}}, z_{\rmn{s}})}
{d_{\rmn{A}} (0, z_{\rmn{s}})}
\end{equation}
is the Einstein angle\footnote{If
the source, lens and observer are colinear, a circular image is be
formed. The angular radius of this circle is the Einstein angle.}
of the singular ($r_{\rmn c} = 0$) lens.
The dynamical normalisation, $\sigma_\infty$, is given
in terms of observables in 
\sect{dynamics}.

\Eq{dimensionless lens equation}
and `a little algebra' (Hinshaw \& Krauss 1987) give
\begin{eqnarray}
\label{equation:cubic}
0 & = & \theta^3 - 2 \beta \theta^2
+ (\beta^2 + 2 \theta_{\rmn{E}} \theta_{\rmn{c}}
- \theta_{\rmn{E}}^2) \theta
- 2 \theta_{\rmn{E}} \theta_{\rmn{c}} \beta
\nonumber \\
& = & \left(\theta - \frac{2}{3}\beta\right)^3
- 3 p \left(\theta - \frac{2}{3}\beta\right) - 2 q,
\end{eqnarray}
where $p = (3 \theta_{\rmn{E}}^2 + \beta^2
- 6 \theta_{\rmn{E}} \theta_{\rmn{c}}) / 9$
and $q = \beta(9 \theta_{\rmn{E}}^2 + 9 \theta_{\rmn{E}} \theta_{\rmn{c}}
- \beta^2) / 27$.
There are three solutions for $\theta$ if
$q^2 < p^3$, two if $q^2 = p^3$ and only one if $q^2 > p^3$.
Note that not all solutions need not correspond to image
positions, as \eq{cubic} is not strictly equivalent to
\eq{dimensionless lens equation}.

If $\theta_{\rmn{c}} > \theta_{\rmn{E}} / 2$, \eq{cubic}
can have only ever have one solution,
as the central surface density of the lens is less than $\Sigma_{\rmn crit}$
[\eq{sigma_crit}].
However, all but the most nearby galaxies 
have $\theta_{\rmn{c}} < \theta_{\rmn{E}} / 2$
and so can form multiple images.
In this case, expanding $p$ and $q$, and disregarding multiple solutions
that do not correspond to images, yields the result
that a source is multiply-imaged is
$\beta \leq \beta_{\rmn crit} = \beta_-$, where
\begin{equation}
\label{equation:beta_crit}
\beta_\pm
= 
\end{equation}
\[
\mbox{}
\left\{ \!\!\!\!
\begin{array}{lll}
\left[\theta_{\rmn{E}}^2
+ 5 \theta_{\rmn{E}} \theta_{\rmn{c}}
- \frac{1}{2}\theta_{\rmn{c}}^2
\pm \frac{1}{2} \sqrt{\theta_{\rmn{c}} (\theta_{\rmn{c}}
+ 4 \theta_{\rmn{E}})^3} \right]^{1/2}, \!\!\!\!
& {\rmn if} \!\!\!\!
& \theta_{\rmn{c}} < \theta_{\rmn{E}} / 2, \\ \\
0, \!\!\!\!
& {\rmn if} \!\!\!\! & \theta_{\rmn{c}} \geq \theta_{\rmn{E}} / 2.
\end{array} \right.
\!\!\!\!\!\!\!\!\!
\]
If $r_{\rmn c} = 0$, the lens is always critical,
and $\beta_{\rmn crit} = \theta_{\rmn E}$.
The cross-section for multiple-imaging is then simply
$\pi \beta_{\rmn crit}^2$, but the image positions and magnifications
are needed for the more realistic lensing calculation
(\sect{prob}).
If $\theta_{\rmn c} \ll \theta_{\rmn E}$
and $\sigma_{||}$ is fixed, $\beta_{\rmn crit}$ decreases slowly with
core radius, as there is little
increase in $\sigma_{\infty}$. 
For larger core radii, however, $\sigma_{\infty}$ 
increases nearly linearly with $r_{\rmn c}$, so, from
\eq{einstein angle}, $\beta_{\rmn crit} \propto \theta_{\rmn E}
\propto r_{\rmn c}^2$.
Hence $\beta_{\rmn crit}$ eventually increases with core radius. 
For large galaxies this only occurs for unrealistically large
values of $r_{\rmn c}$, 
but it can come into effect for very small galaxies if 
the fibre integration area is small
(\sect{tau}).

If $(0\leq)$ $\beta \leq \beta_{\rmn crit}$, the source is lensed,
and the three images are located at
\begin{equation}
\label{equation:lensed image positions}
\theta_i (\beta) = \frac{2}{3} \beta
+ 2 \sqrt{p} \cos\left[
\frac{1}{3} \arccos\left(\frac{q}{\sqrt{p^3}} \right)
+ \frac{2 i \pi}{3}
\right],
\end{equation}
where $i = $1, 2 or 3, and two of the images are coincident if 
$\beta = \beta_{\rmn crit}$. 
Note that this formula breaks down if the lens is singular,
in which case only two images are formed.
They are located at
$\theta_1 = \beta + \theta_{\rmn E}$ and
$\theta_2 = \beta - \theta_{\rmn E}$.
To within 10 per cent, the angular separation between the outer pair
of images is independent of source position, and is given by
(Hinshaw \& Krauss 1987)
\begin{equation}
\label{equation:del_th_core}
\Delta \theta \simeq
        \left\{
        \begin{array}{lll}
        0, & {\rmn if} & \theta_{\rmn c} > \theta_{\rmn E} / 2, \\
        & & \\
        \theta_{\rmn E}
        \sqrt{1 - 2 \theta_{\rmn c} / \theta_{\rmn E}},
        & {\rm if} & \theta_{\rmn c} \leq \theta_{\rmn E} / 2.
        \end{array}
        \right. 
\label{equation:delta_th}
\end{equation}
This expression becomes exact in the singular case.

If $\beta > \beta_{\rmn crit}$,
the source is unlensed, and there is only one image.
For the range $\beta_{\rmn crit} = \beta_- < \beta \leq \beta_+$
the single image position is
\[
\theta (\beta) = \frac{2}{3} \beta +
\left( q - \sqrt{q^2 - p^3} \right)^{1/3}
+ p \left( q - \sqrt{q^2 - p^3} \right)^{-1/3}.
\]
\begin{equation}
\end{equation}
If $\beta > \beta_+$, \eq{cubic}
has three solutions, two of which do not represent image positions.
The one solution that does is
\begin{equation}
\theta (\beta) =
\frac{2}{3} \beta
+ 2 \sqrt{p} \cos\left[
\frac{1}{3} \arccos\left(\frac{q}{\sqrt{p^3}} \right)\right].
\end{equation}
Again the singular case is much simpler: if 
$\beta > \theta_{\rmn E}$, the single image
is located at $\theta = \beta + \theta_{\rmn E}$.

The lens mapping (from $\theta$ to $\beta$, or image to source)
changes the area subtended by an object; this manifests itself
as an increase or decrease in flux, as surface brightness is 
conserved by gravitational lensing. 
The area change
is given by the Jacobian of the mapping,
and so the magnification is the reciprocal of this.
For spherically-symmetric lenses (See Schneider \etal\ 1992.)
the magnification of a point source
is given by 
\begin{equation}
\mu(\theta) = 
\left| \frac{\beta}{\theta} \frac{\partial\beta}{\partial\theta} \right|.
\end{equation}
\Eq{dimensionless lens equation} then gives
\begin{eqnarray}
\label{equation:image magnification}
\mu(\theta) & = &
\left| \frac{\theta^2}
{\theta^2 + \theta_{\rmn{E}} \theta_{\rmn{c}} - \theta_{\rmn{E}}
\sqrt{\theta^2 + \theta_{\rmn{c}}^2}} \right| 
\nonumber \\
& \times & 
\left|
\frac{\theta^2 \sqrt{\theta^2 + \theta_{\rmn{c}}^2}}
{\theta_{\rmn{E}} \theta_{\rmn{c}}^2 + \left(\theta^2 - \theta_{\rmn{E}}
\theta_{\rmn{c}}\right)
\sqrt{\theta^2 + \theta_{\rmn{c}}^2}}
\right|.
\end{eqnarray}
This expression breaks down for for small $\theta$
if $\theta_{\rmn{c}} < \theta_{\rmn{E}} / 2$, but a Taylor expansion
about $\theta = 0$ gives
\begin{equation}
\mu(\theta) \simeq \frac{4 \theta_{\rmn{c}}^2 / \theta_{\rmn{E}}^2}
{(2 \theta_{\rmn{c}} / \theta_{\rmn{E}} - 1)^2 +
(2 \theta_{\rmn{c}} / \theta_{\rmn{E}} - 1) \theta^2 / \theta_{\rmn{c}}^2},
\end{equation}
for $\theta \ll \theta_{\rmn{E}}$.
If $r_{\rmn c} = 0$, \eq{image magnification} 
reduces to $\mu(\theta) = |\theta / (|\theta| - \theta_{\rmn E})|$.

The total magnification of a source is simply the sum of the
magnification of its images, and is thus
\begin{equation}
\label{equation:mu_tot}
\mu_{\rmn tot} (\beta) = \sum_i \mu\left[ \theta_i(\beta) \right].
\end{equation}
The flux ratio is less well-defined, due to potential presence 
of three images and observational effects (\eg\ Kassiola \&
Kovner 1993). For three image configurations it is taken to
be the flux ratio of the two brightest images.

\subsection{Optical depth}
\label{section:tau} 

\begin{figure}
\includegraphics{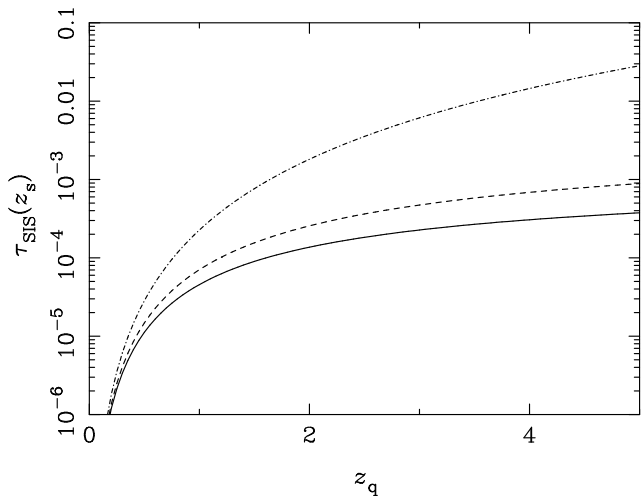}
\vspace{\singlefigureheight}
\caption{The `standard' lensing optical depth, $\tau_{\rmn SIS}$,
as a function of source redshift, $z_{\rmn s}$.
This results from a population of singular isothermal spheres with
$\sigma_\infty = \sigma_{||}$, and the default parameters
described in \sect{models}.
Three cosmological models are shown:
$\Omega_{\rm m_0} = 1$ and $\Omega_{\Lambda_0} = 0$ (solid line);
$\Omega_{\rm m_0} = 0$ and $\Omega_{\Lambda_0} = 0$ (dashed line);
and
$\Omega_{\rm m_0} = 0$ and $\Omega_{\Lambda_0} = 1$ (dot-dashed line).}
\label{figure:tau_z}
\end{figure}

The lensing optical depth, $\tau$, as 
introduced by Turner \etal\ (1984),
is the fraction of the source plane within which the
lens equation has multiple solutions.
It is a useful estimate of the lensing probability
that is independent of observational restrictions and the source 
luminosity function. 
In the filled-beam approximation, the contribution to the
optical depth by any one lens galaxy, $\tau_{\rmn g}$,
is the fraction of the sky covered by its cross-section.
Hence $\tau_{\rmn g} = \pi \beta_{\rmn crit}^2 / (4 \pi)$,
with $\beta_{\rmn crit}$ given in \eq{beta_crit}.
The optical depth 
is given by integrating $\tau_{\rmn g}$
over the population of deflectors, 
under the assumption 
that the individual cross-sections do not 
overlap.
For a source at redshift $z_{\rmn s}$,
\begin{equation}
\tau(z_{\rmn s}) = \int_0^{z_{\rmn s}} \int_0^\infty
\frac{{\rmn d}^2 N_{\rmn g}}{{\rmn d}z_{\rmn g} \, {\rmn d}\sigma_{||}}
\tau_{\rmn g} \,
{\rmn d}\sigma_{||} \, {\rmn d}z_{\rmn g}.
\end{equation}
If the lenses are singular and $\sigma_\infty = \sigma_{||}$ is 
assumed, $\tau_{\rmn g} = \theta_{\rmn E}^2 / 4$ and 
the integrals can be performed analytically
(\eg\ Turner \etal\ 1984; Schneider \etal\ 1992; 
Krauss \& White 1992; Kochanek 1993) to give
\begin{equation}
\tau_{\rmn SIS}(z_{\rmn s})
 = 16 \pi^3 n_* \left( \frac{c}{H_0} \right)^3 
\left(\frac{\sigma_{||}}{c}\right)^4
\Gamma \left(1 + \alpha + \frac{4}{\gamma} \right)
\end{equation}
\[
\mbox{} 
 \times \left\{
\begin{array}{lll}
\frac{4}{15} \left(1 - \frac{1}{\sqrt{z_{\rmn s} + 1}} \right)^3,
& \!\!\!\!\!\!\!\!\!\!\!\!\!\!\!\!\!\!\!\!\!\!\!\!\!\!\!\!\!\!
\!\!\!\!\!\!\!\!\!\!\!\!\!\!\!\!\!\!\!\!\!\!\!\!\!\!\!\!\!\!
\!\!\!\!\!\!\!\!\!
{\rmn if} & 
\Omega_{\rmn m_0} = 1 {\,\, \rmn and \,\,} \Omega_{\Lambda_0} = 0, \\ \\
\frac{2 \left(z_{\rmn s}^4 + 4 z_{\rmn s}^3 + 10 z_{\rmn s}^2 + 
	12 z_{\rmn s} + 6 \right) \ln(z_{\rmn s} + 1)
	- 3 z_{\rmn s} (z_{\rmn s} + 2) 
        \left(z_{\rmn s}^2 + 2 z_{\rmn s} + 2 \right)}
	{8 z_{\rmn s}^2 (z_{\rmn s} + 2)^2}, \\
& \!\!\!\!\!\!\!\!\!\!\!\!\!\!\!\!\!\!\!\!\!\!\!\!\!\!\!\!\!\!
\!\!\!\!\!\!\!\!\!\!\!\!\!\!\!\!\!\!\!\!\!\!\!\!\!\!\!\!\!\!
\!\!\!\!\!\!\!\!\!
{\rmn if} &
\Omega_{\rmn m_0} = 0 {\,\, \rmn and \,\,} \Omega_{\Lambda_0} = 0, \\ \\
\frac{1}{30} z_{\rmn s}^3, 
& \!\!\!\!\!\!\!\!\!\!\!\!\!\!\!\!\!\!\!\!\!\!\!\!\!\!\!\!\!\!
\!\!\!\!\!\!\!\!\!\!\!\!\!\!\!\!\!\!\!\!\!\!\!\!\!\!\!\!\!\!
\!\!\!\!\!\!\!\!\!
{\rmn if} &
\Omega_{\rmn m_0} = 0 {\,\, \rmn and \,\,} \Omega_{\Lambda_0} = 1,
\end{array}
\right.
\]
where \eqs{schechter}, (\ref{equation:d2n_gal}), 
(\ref{equation:dv_0}), (\ref{equation:d_a_core}) and 
(\ref{equation:einstein angle}) 
have been used.
The prefactor is the same as the $F$-parameter
introduced by Turner \etal\ (1984) to scale the lensing
effectiveness of a galaxy population; for the 
elliptical galaxy population described in 
\sect{population}, $F = 0.008 \pm 0.002$.
\Fig{tau_z} shows $\tau_{\rmn SIS}(z_{\rmn s})$ 
for the three cosmological models, showing the expected 
increase with redshift as well as the strong dependence on 
$\Omega_{\Lambda_0}$. 

\begin{figure}
\includegraphics{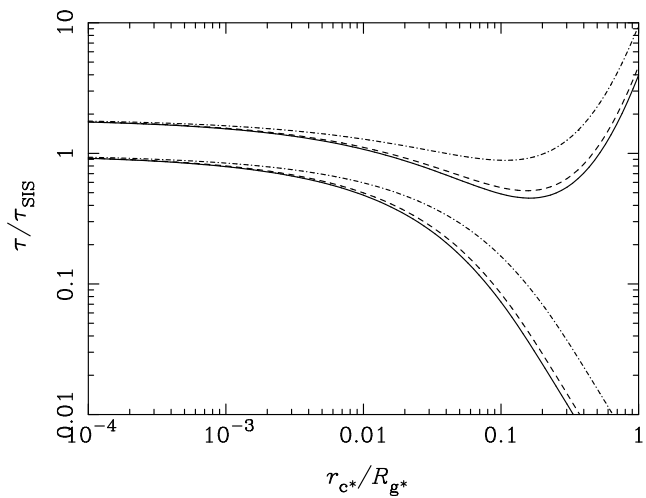}
\vspace{\singlefigureheight}
\caption{The lensing optical depth, $\tau$, of a source at
$z_{\rmn s} = 2$, scaled by the `standard' lensing optical depth for
a population of singular isothermal spheres with $\sigma_\infty = \sigma_{||}$.
Three cosmological models are shown:
$\Omega_{\rm m_0} = 1$ and $\Omega_{\Lambda_0} = 0$ (solid lines);
$\Omega_{\rm m_0} = 0$ and $\Omega_{\Lambda_0} = 0$ (dashed lines);
and
$\Omega_{\rm m_0} = 0$ and $\Omega_{\Lambda_0} = 1$ (dot-dashed lines).
The lower set of lines are for $\sigma_\infty = \sigma_{||}$.
The values of the other parameters are taken to be:
$\beta_\sigma = 0$;
$u_{\rmn c} = 4$;
$u_{\rmn g} = 4$;
and $R_{\rmn f} = R_{\rmn g}$.}
\label{figure:tau}
\end{figure}

The optical depth (normalised to the above analytic cases) is
shown as a function of core radius in \fig{tau}.
Independent of cosmology, the linear conversion
$\sigma_\infty \simeq 1.1 \sigma_{||}$ results in
$\tau / \tau_{\rmn SIS} \simeq 1.5$;
if the conversion is $\sigma_\infty \simeq 1.2 \sigma_{||}$,
then
$\tau / \tau_{\rmn SIS} \simeq 2.1$, and the
optical depth is doubled.
The lower set of lines in \fig{tau} show the 
marked decrease in $\tau$ with core radius that was first 
demonstrated by Hinshaw \& Krauss (1987).
The upper set of lines use the normalisation described in 
\Sect{models}, and show not only the higher optical depth
for singular models, but a very different dependence on core radius. 
For large $r_{\rmn c}$, the effects of the normalisation becomes
very important, 
and `small' galaxies (with $\sigma_{||} \ll \sigma_*$) dominate
the optical depth as their measured dispersion 
is purely that in the under-dense core. The up-turn in $\tau$ is not 
necessarily realistic, and occurs for core radii that are 
greater than observed -- Kochanek (1996b)
found $r_{\rmn c} \leq 0.08 R_{\rmn g}$
at 95 per cent confidence.
For moderate core radii the normalisation is still important --
in the $\Omega_{\rm m_0} = 0$ and $\Omega_{\Lambda_0} = 1$ 
cosmology, $\tau / \tau_{\rmn SIS} \ga 1$ for all values of 
$r_{\rmn c}$. It is also potentially important that a
finite core radius actually enhances the cosmological dependence
of $\tau$, as most of the lenses are at high-redshift if 
$\Omega_{\Lambda_0} > 0$ (Kochanek 1992).

The dependence of $\tau$ on the various galaxy population parameters
is shown in Mortlock (1999), but there is little difference 
between the resultant plots and those shown in \fig{p_q}
for the quasar lensing probability, and so the optical depth 
plots are omitted here.

\subsection{Quasar lensing probability}
\label{section:prob}

The optical depth is a useful measure of lensing likelihood, 
but it cannot be directly compared to measured lensing frequencies.
Hence the probability that a quasar is observed to be lensed,
$p_{\rmn q}$, must be calculated.
In the case of the generic lens survey, all the quasars in a 
parent survey are re-examined for secondary lensed images.
Under these conditions, $p_{\rmn q}$ is the
fraction of all redshift $z_{\rmn q}$ quasars of magnitude $m_{\rmn q}$ 
(as measured in the parent survey)
that would be revealed as lenses if examined 
with the resolution and sensitivity of the secondary search.
The resolution limit implies a minimum image separation, 
$\Delta \theta_{\rm min}$, and the depth of the 
follow-up observations 
leads to a maximum magnitude difference between the 
primary and secondary images, $\Delta m_{\rm max}$. 
In general there is also a maximum image separation, 
$\Delta \theta_{\rm max}$, defined by the extent of the 
search for companion images.

Most lensed sources are magnified by a
factor of 2 or more, so the number of lenses at magnitude $m_{\rmn q}$
is determined by the quasar number counts at least one magnitude 
fainter. The quasar luminosity function is so steep that this
magnification bias (Turner 1980)
can double or triple the estimated lensing probability.
The quasar luminosity function is taken to satisfy
\begin{equation}
\frac{{\rmn d}^2N_{\rmn q}}{{\rmn d} z \,{\rmn d} m} \propto
\frac{1}{
{10^{-\alpha_{\rmn q}(m - m_{\rmn q0})}
- 10^{-\beta_{\rmn q}(m - m_{\rmn q0})}}}
\end{equation}
at all redshifts,
where $m_{\rmn q0} = 19.0 \pm 0.2$ is the magnitude of the break in
the number counts,
$\alpha_{\rmn q} = 0.9 \pm 0.1$ is the bright-end slope,
$\beta_{\rmn q} = 0.3 \pm 0.1$ is the faint-end slope
(Boyle, Shanks \& Peterson 1988; 
Kochanek 1996b), and the normalisation unimportant.

The probability a quasar is lensed 
by a particular galaxy
is given by integrating over the source position as
(\eg\ Kochanek 1995, 1996b)
\begin{equation}
\label{equation:p_qg_core}
p_{\rmn q,g} = 
\frac{
\int_0^{\beta_{\rmn crit}}
2 \pi \beta S(\beta)
\left. \frac{{\rmn d}^2N_{\rmn q}}{{\rmn d} z_{\rmn q}\, {\rmn d} m}
\right|_{m = m_{\rmn q} + 5/2\, \log[ \mu_{\rmn tot} (\beta)]}
\, {\rmn d} \beta}
{4 \pi \, \frac{{\rmn d}^2N_{\rmn q}}
{{\rmn d} z_{\rmn q}\, {\rmn d} m_{\rmn q}}},
\end{equation}
where $S(\beta)$ is the selection function.
It can be approximated by
\begin{eqnarray}
S(\beta) & = & H \! \left[\Delta m_{\rmn max} - \Delta m(\beta)\right]
\nonumber \\
& \times &
H \! \left[\Delta \theta (\beta) - \Delta \theta_{\rmn min} \right]\,
H \! \left[\Delta \theta_{\rmn max} - \Delta \theta (\beta) \right],
\end{eqnarray}
where $H(x)$ is the Heavyside step function.
Hence $S(\beta) = 1$ 
if the images of a source at position $\beta$ satisfy the resolution
and sensitivity limits, and is zero otherwise.
The 
use of $\mu_{\rmn tot} (\beta)$ [given in \eq{mu_tot}]
in the calculation of the
magnification bias implies that the all the images of the 
source were unresolved in the parent survey\footnote{This is almost always
a valid assumption in the case of
galactic lenses, but can result in a serious over-estimate
of the magnification bias for more massive deflectors 
(Mortlock \& Webster 2000a).},
and 
the pre-factor converts the expression from the surface
density of lenses to the fraction of quasars which are lensed.

\begin{figure}
\includegraphics{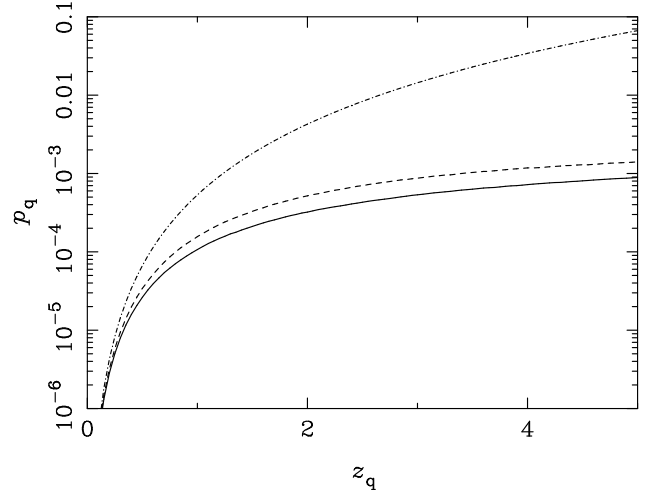}
\vspace{\singlefigureheight}
\caption{The probability that 
a $m_{\rmn q} = 19$ quasar at redshift $z_{\rmn q}$
is observed to be multiply-imaged, assuming that all secondary images 
brighter than $m = 22$ and
separated by $\geq 1$ arcsec from the primary 
are found.
This results from a population of singular isothermal spheres with
$\sigma_\infty = \sigma_{||}$, and the default population parameters
described in \sect{models}.
Three cosmological models are shown:
$\Omega_{\rm m_0} = 1$ and $\Omega_{\Lambda_0} = 0$ (solid line);
$\Omega_{\rm m_0} = 0$ and $\Omega_{\Lambda_0} = 0$ (dashed line);
and
$\Omega_{\rm m_0} = 0$ and $\Omega_{\Lambda_0} = 1$ (dot-dashed line).}
\label{figure:p_q_z}
\end{figure}

Integrating $p_{\rmn q,g}$ over the deflector population
yields (\cf\ Kochanek 1996b)
\begin{equation}
\label{equation:p_q_core}
p_{\rmn q} = \int_0^{z_{\rmn q}} \int_0^\infty
\frac{{\rmn d} V_0}{{\rmn d} z_{\rmn g}}
\frac{{\rmn d} n_{\rmn g}}{{\rmn d} \sigma_{||}}
p_{\rmn q,g} \,
{\rmn d}\sigma_{||} \, {\rmn d}z_{\rmn g}.
\label{equation:p_q}
\end{equation}
This is somewhat simpler for the standard singular lens models (\eg\ 
Kochanek 1993), but no closed form expression for $p_{\rmn q}$ 
is available unless a simpler form of the 
quasar luminosity function is used as well.

\Fig{p_q_z} shows the probability that an $m_{\rmn q} = 19$
quasar is observed to be lensed by an elliptical galaxy as 
a function of source redshift.
The survey parameters chosen are $\Delta \theta_{\rmn min}
= 1$ arcsec,
$\Delta \theta_{\rm max} \rightarrow \infty$\footnote{Galactic
lenses are incapable of producing image separations of more
than a few arcsec, so $\Delta \theta_{\rmn max}$ is usually
unimportant.}
and $\Delta m_{\rmn max} = 3$. These 
values affect the overall probability considerably (\eg\
Schneider \etal\ 1992), but do not strongly influence 
the relative dependence on the other parameters.
The lens probability is consistently a factor of $\sim2$ higher 
than the optical depth shown in \fig{tau_z}.
Although some lenses are lost due to the angular separation cut-off,
many more are magnified into the survey. As with the optical 
depth, $p_{\rmn q}$ is strongly-dependent on the cosmological
model -- the similarity of \figs{tau_z} and 
\ref{figure:p_q_z} demonstrates the generic nature of the
cosmological dependence.

\begin{figure*}
\includegraphics{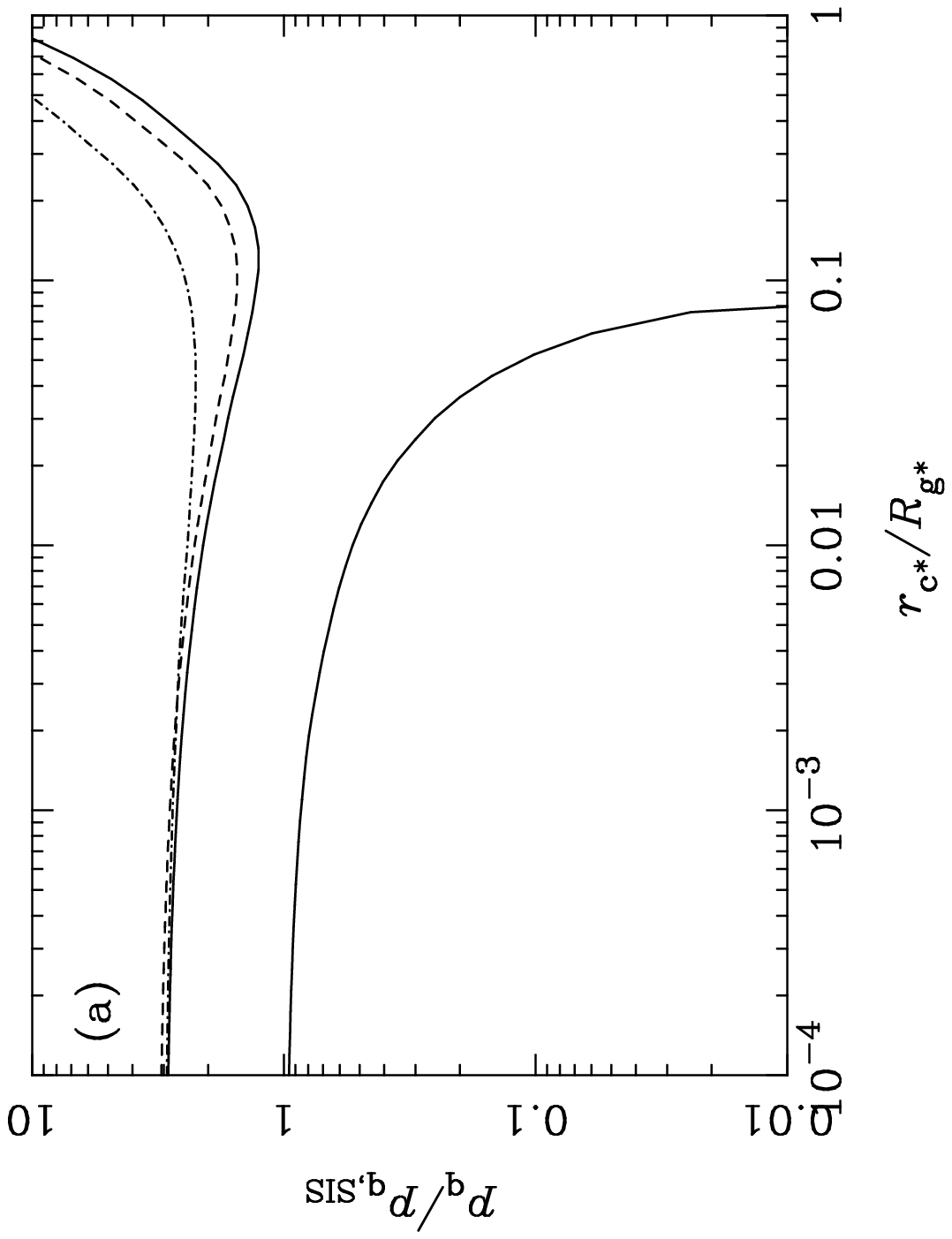}
\includegraphics{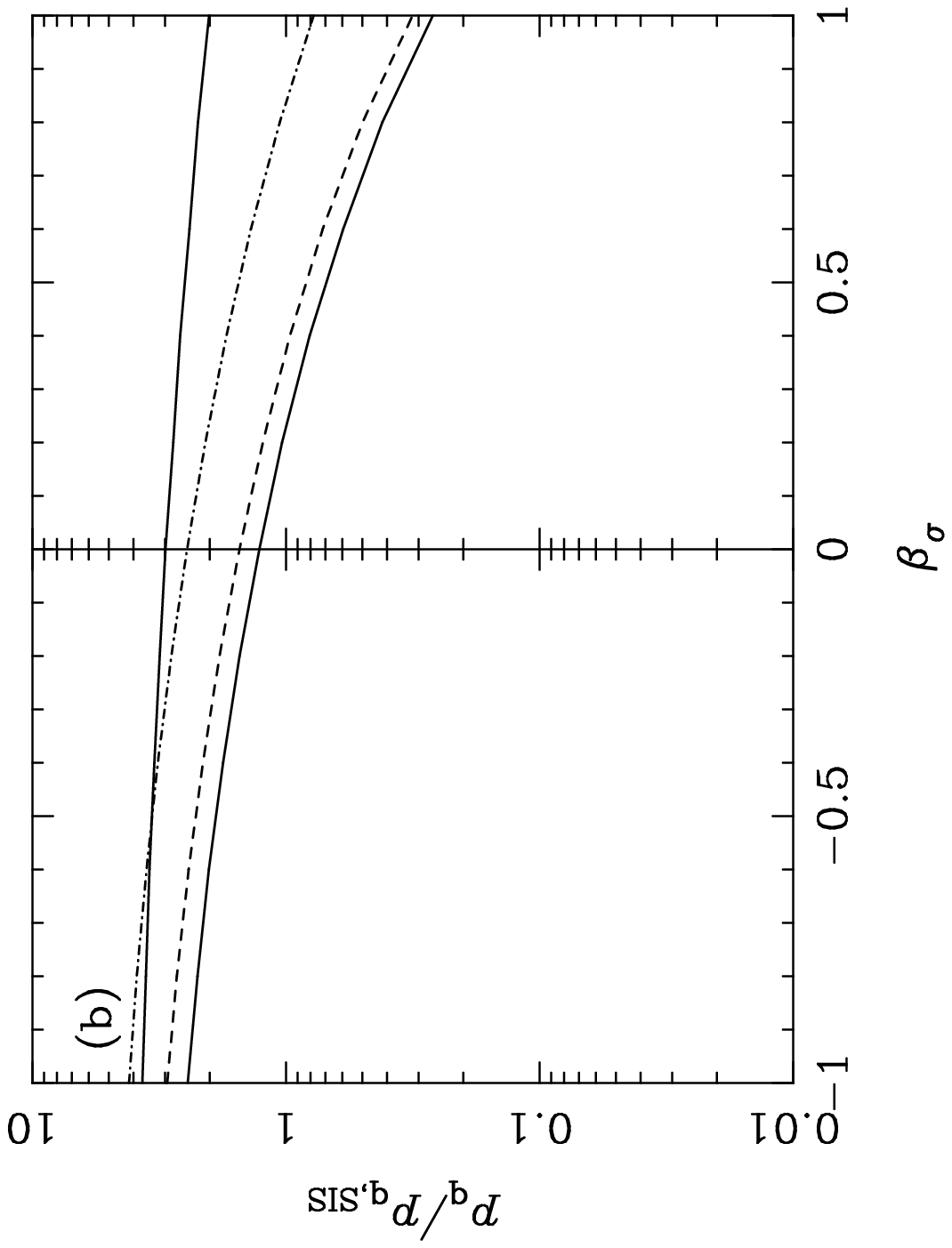}
\includegraphics{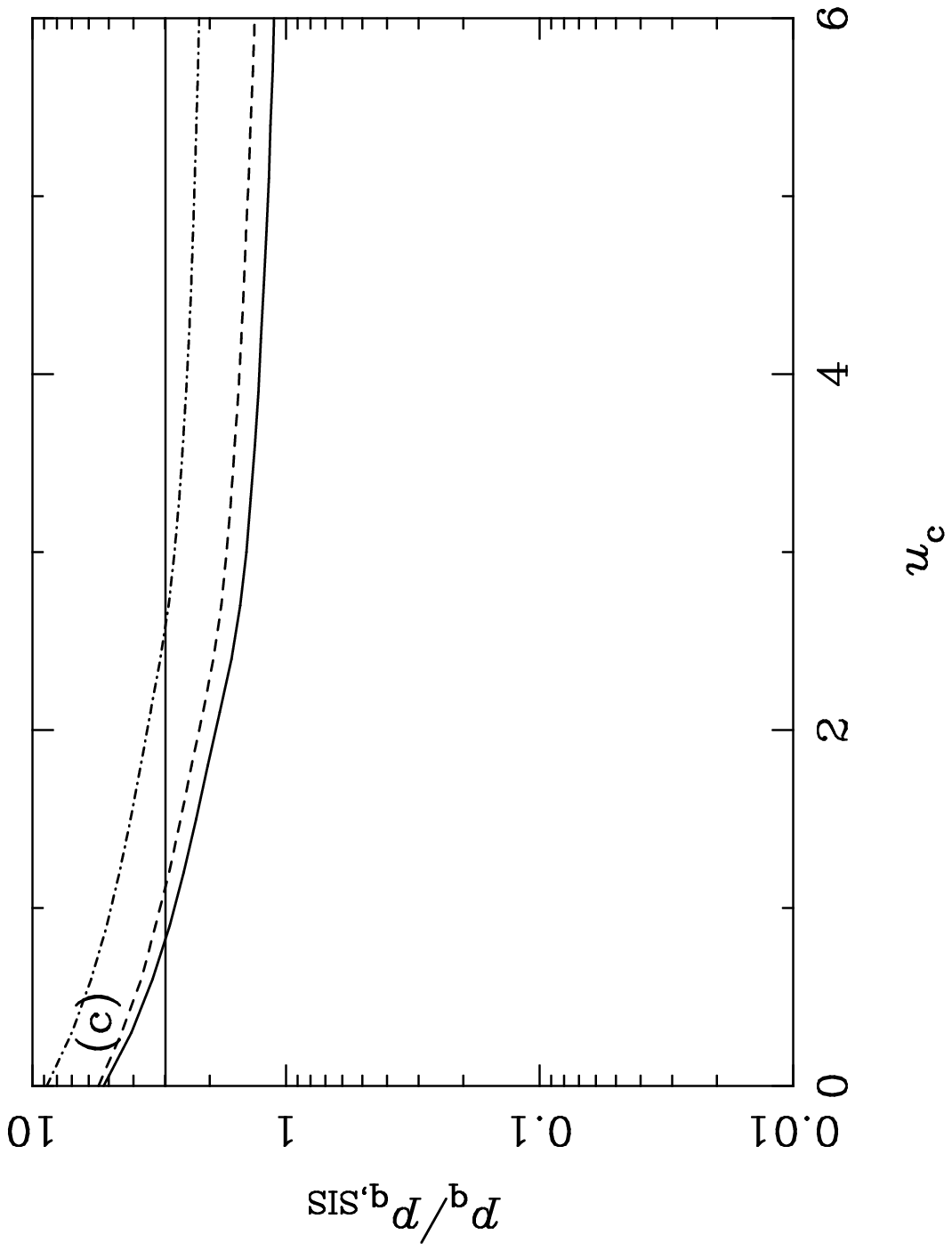}
\includegraphics{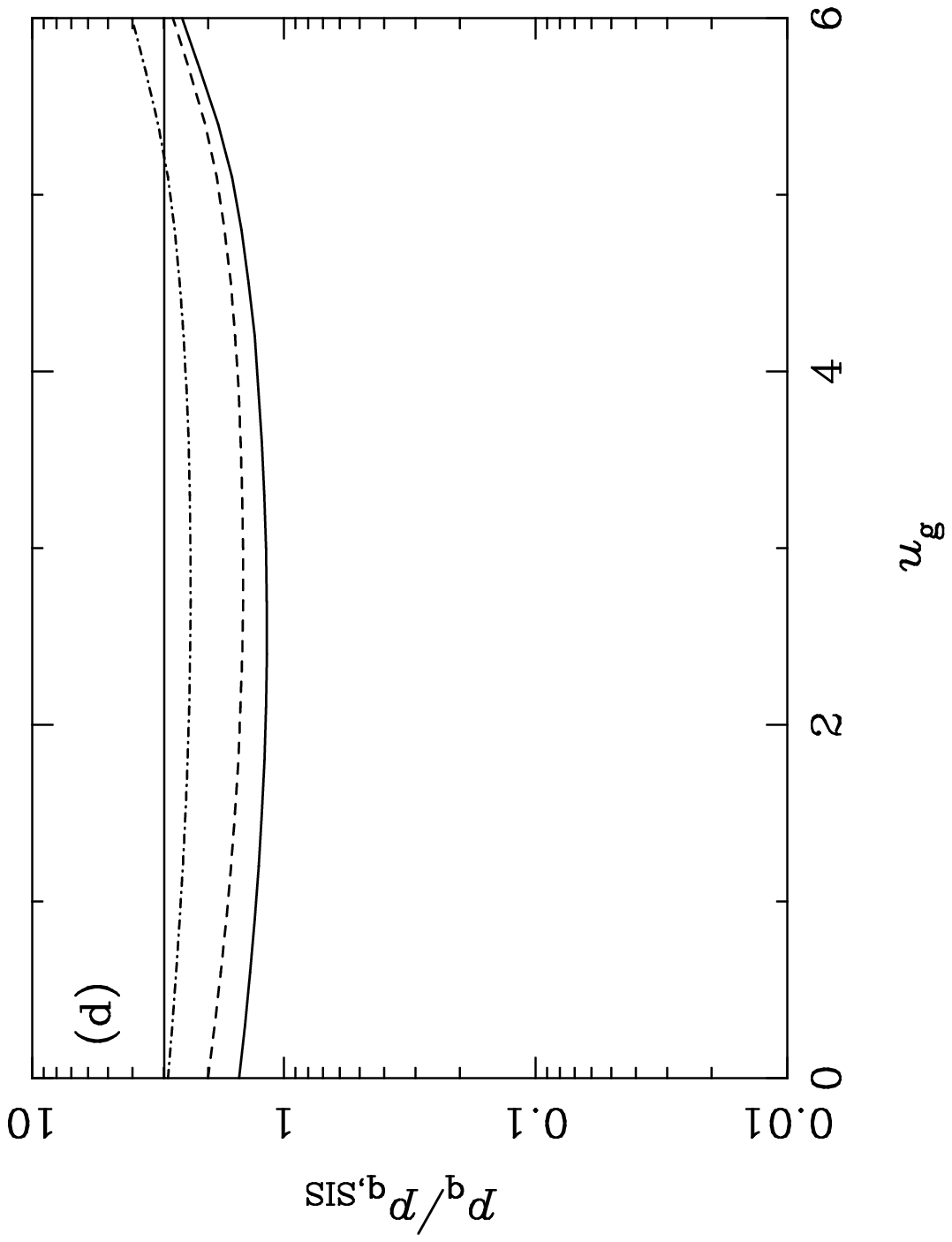}
\includegraphics{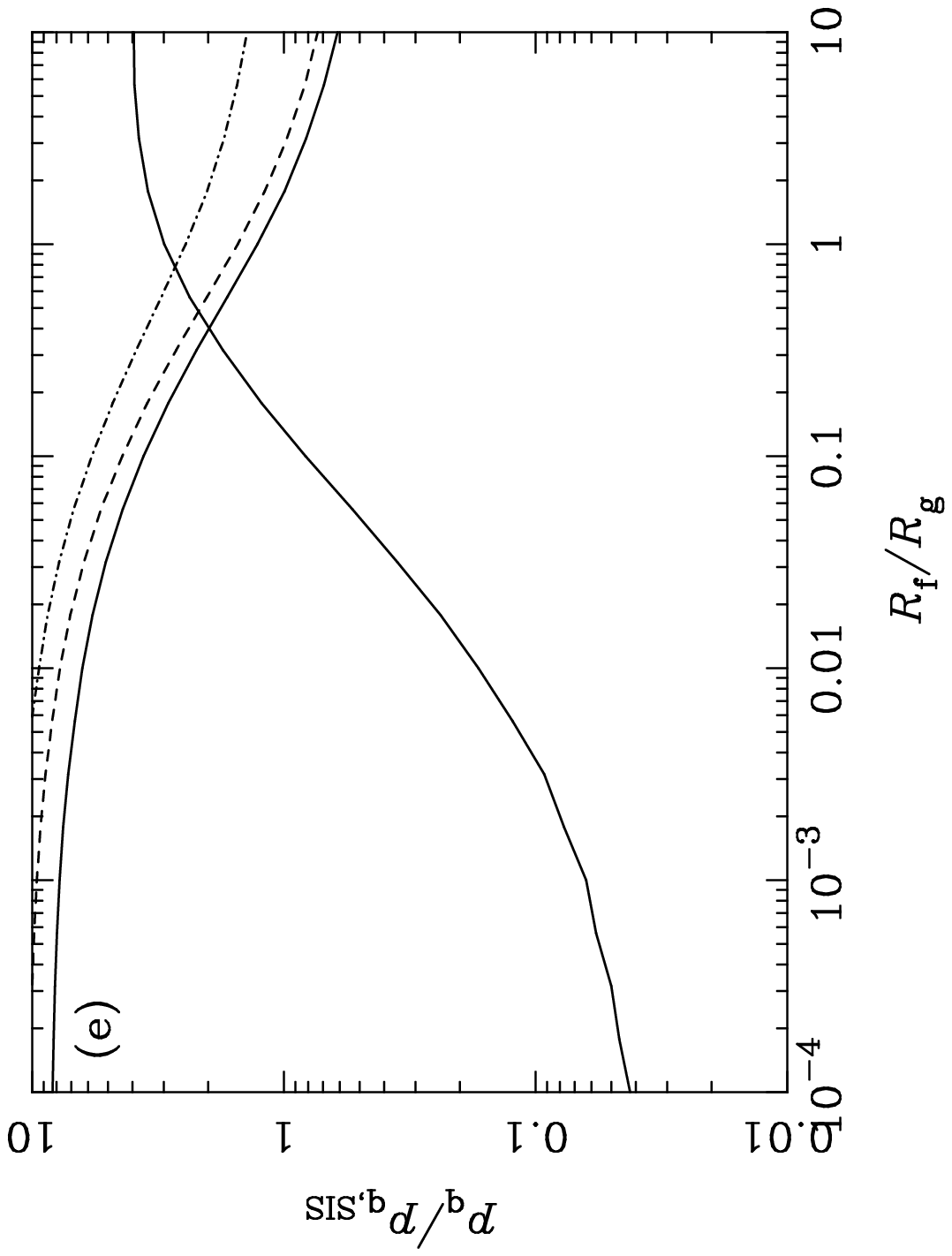}
\includegraphics{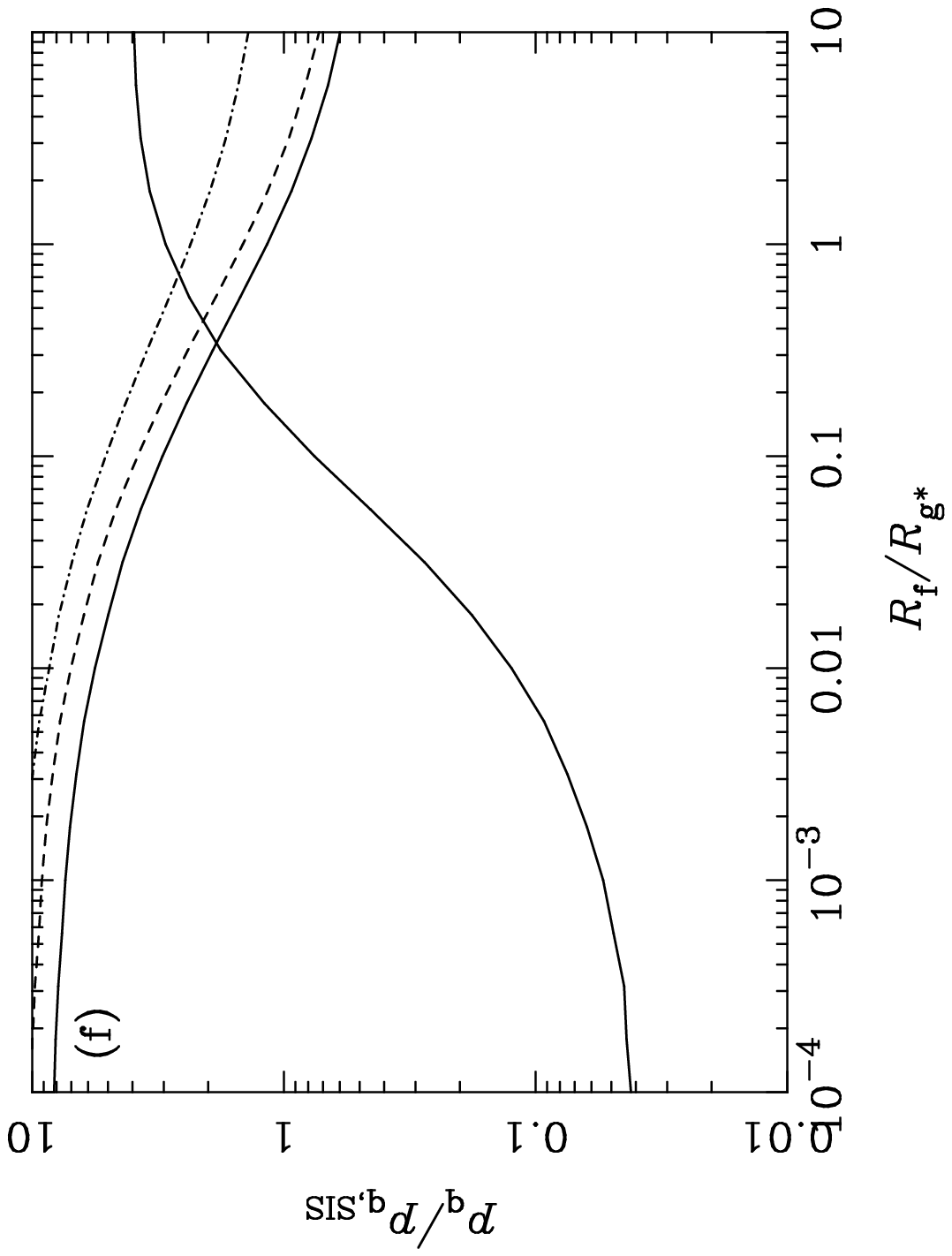}
\vspace{\triplefigureheight}
\caption{The probability that a $m_{\rmn q} = 19$, $z_{\rmn q} = 2$ quasar is
lensed with $\Delta \theta \geq 1$ arcsec.
It is scaled by the `standard' lensing probability of
a population of singular lenses with $\sigma_\infty = \sigma_{||}$.
Three cosmological models are shown in each panel:
$\Omega_{\rm m_0} = 1$ and $\Omega_{\Lambda_0} = 0$ (solid lines);
$\Omega_{\rm m_0} = 0$ and $\Omega_{\Lambda_0} = 0$ (dashed lines);
and
$\Omega_{\rm m_0} = 0$ and $\Omega_{\Lambda_0} = 1$ (dot-dashed lines).
The lower line in (a) is for $\sigma_\infty = \sigma_{||}$;
the two sets of lines in panels (b), (c), (d), (e) and (f) represent
$r_{\rmn c*} = 0$ (single line as there is no cosmological dependence) and
$r_{\rmn c*} = 0.1 R_{\rmn g*}$ (three distinct lines).
The different panels show the dependence of $p_{\rmn q}$ on:
scale core radius, $r_{\rmn c *}$, in (a);
velocity anisotropy, $\beta_\sigma$, in (b);
core radius scaling exponent, $u_{\rmn c}$, in (c);
effective radius scaling, $u_{\rmn g}$, in (d);
and the aperture size, $R_{\rmn f}$, in (e) and (f).
In (e) $R_{\rmn f} = R_{\rmn g}$, scaling with galaxy size;
in (f) $R_{\rmn f} = R_{\rmn g*}$ and is constant.
The defaults values used are:
$\beta_\sigma = 0$;
$u_{\rmn c} = 4$;
$u_{\rmn g} = 4$;
and $R_{\rmn f} = R_{\rmn g}$.}
\label{figure:p_q}
\end{figure*}

\Fig{p_q} shows the dependence of $p_{\rmn q}$ on
various galaxy population parameters.
Firstly, \fig{p_q} (a) shows the variation of 
lensing probability with core radius, in analogy with 
the optical depth dependence shown in \fig{tau}. 
In the case of the unnormalised model (the single solid line),
the high-$r_{\rmn c*}$ cut-off is even more pronounced than 
for $\tau$, despite the effects of the magnification bias.
This occurs as the image separation drops
below $\Delta \theta_{\rmn min}$ quite quickly.
For the correctly normalised models (the set of three lines),
the enhancement in $p_{\rmn q}$ is even greater than that for
the optical depth (\fig{tau}). 
This comes about both due to the magnification bias
and the increased deflection angles of the smaller deflectors.
The increase in $\Delta \theta$ for a given $\sigma_{||}$ is 
another important effect of the dynamical normalisation.

\Fig{p_q} (b) shows that the velocity anisotropy,
$\beta_\sigma$ is not as important as the core radius in lens
statistics. Whilst the difference between $\beta_\sigma = - 1$
and $\beta_\sigma = 1$ can be a factor of several,
most results suggest that $\beta_\sigma \simeq 0$ for
ellipticals, as discussed in \sect{dynamics}.
The lensing probability decreases with $\beta_\sigma$ as
the observed dispersion within a fixed mass distribution increases
with the dominance of radial orbits.

The variation of core radius and effective radius with velocity dispersion
is usually unimportant in lensing calculations,
as illustrated by the
flat parts of the curves in \fig{p_q} (c) and (d)
as well as Krauss \& White (1992).
However, in some situations,
the values of $u_{\rmn c}$ and $u_{\rmn g}$
[as defined in \eqs{c scaling} and (\ref{equation:g scaling}),
respectively] can be important (\eg\ Kochanek 1991).
The sharp increases in $p_{\rmn q}$ seen in \fig{p_q} (c) and (d)
occur when $\mbox{$|u_{\rmn g} - u_{\rmn c}|$} \ga 3$, and the smaller galaxies
have core radii comparable to both their effective radius and 
to the scale over which the dispersion is measured.
The massive increase in $p_{\rmn q}$ is probably unrealistic, but a weaker
form of the effect will occur.

The greatest assumption in these calculations is
involved with the choice of $R_{\rmn f}$, the scale over
which the line-of-sight velocity dispersion is measured.
\Fig{p_q} (e) and (f) show $p_{\rmn q}/p_{\rmn q,SIS}$
as a function of the typical scale of $R_{\rmn f}$; in
the former panel it scales with the effective radius of the
galaxies, whereas in the latter it does not vary with the properties of 
the galaxy in question.
If ellipticals are singular, a given observed dispersion
results in a higher optical depth if $R_{\rmn f} \ga R_{\rmn g}$,
as the orbital speeds are much lower in the outer regions.
Conversely, if ellipticals have significant core radii,
the optical depth is highest for small apertures, in which case
the measured dispersion is only a fraction of the
dynamical normalisation. These arguments are true irrespective of
how $R_{\rmn f}$ scales with $R_{\rmn g}$, as
can be seen from the similarity of panels (e) and (f) in \fig{p_q}.

\subsubsection{Cosmological implications}

\begin{figure}
\includegraphics{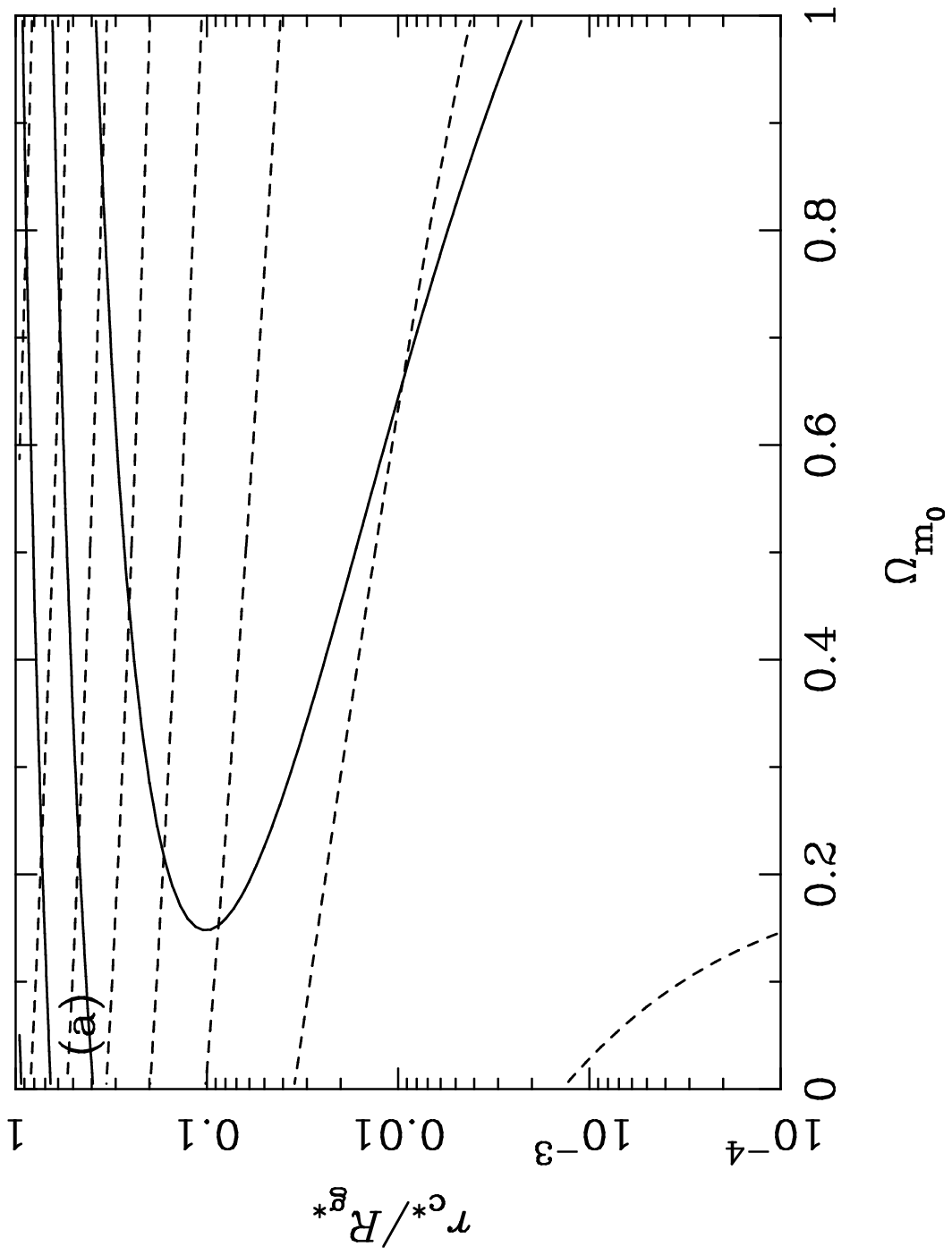}
\includegraphics{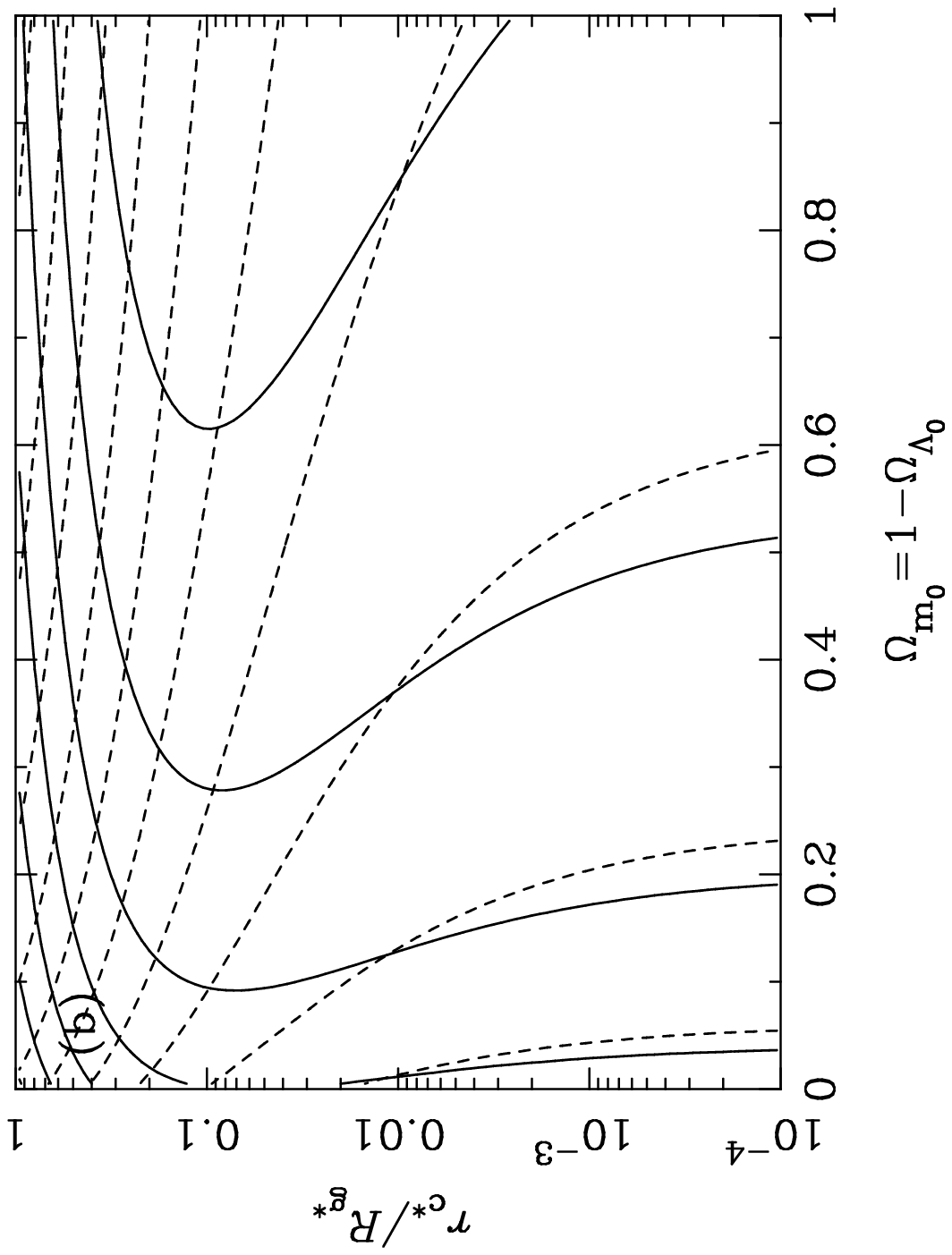}
\vspace{\doublefigureheight}
\caption{Contour plots showing the probability that a
$m_{\rmn q} = 19$, $z_{\rmn q} = 2$ quasar is lensed
with $\Delta \theta \geq 1$ arcsec,
as a function of $\Omega_{{\rmn m}_0}$ 
and the canonical core radius, $r_{\rmn c*}$. 
In (a) there is no cosmological constant (\ie\ $\Omega_{\Lambda_0} = 0$);
in (b) the universe is spatially flat (\ie\ 
$\Omega_{{\rmn m}_0} + \Omega_{\Lambda_0} = 1$).
The contours are spaced logarithmically -- three 
per decade.
In (a) those for the unnormalised lens model (dashed lines)
range from $p_{\rmn q} = 2 \times 10^{-6}$
to $p_{\rmn q} = 5 \times 10^{-4}$;
those for the correctly-normalised model (solid lines)
range from $p_{\rmn q} = 5 \times 10^{-4}$
to $p_{\rmn q} = 5 \times 10^{-3}$.
In (b) those for the unnormalised lens model (dashed lines)
range from $p_{\rmn q} = 2 \times 10^{-6}$
to $p_{\rmn q} = 2 \times 10^{-3}$;
those for the correctly-normalised model (solid lines)
range from $p_{\rmn q} = 5 \times 10^{-4}$
to $p_{\rmn q} = 5 \times 10^{-2}$.}
\label{figure:omega_core}
\end{figure}

The above formulation for $p_{\rmn q}$ was extended to
arbitrary cosmologies, the relevant distance and 
volume element formul\ae\ for which are given in 
\eg\ Carroll \etal\ (1992).
\Fig{omega_core} shows the interdependence of
$p_{\rmn q}$ on the cosmological model and $r_{\rmn c*}$,
both with and without the self-consistent dynamical normalisation.
In the models with $\Omega_{\Lambda_0} = 0$ shown in (a),
the core radius is considerably more important than the
value of $\Omega_{\rm m_0}$, as expected. 
Further, the difference between the normalised
and unnormalised models is greater than that between
any $\Omega_{\Lambda_0} = 0$ cosmological model.
The spatially-flat models 
shown in (b) have a much stronger cosmological dependence,
but even then the core radius becomes more important as it
approaches the effective radius of the lens galaxies.
For more realistic values of $r_{\rmn c*}$, a slightly larger
cosmological constant is permitted than if $r_{\rmn c*} = 0$,
but not relative to the $\sigma_\infty = \sigma_{||}$ models.
For instance, if a given data-set implied an upper limit
of $\Omega_{\Lambda_0} \simeq 0.7$ for 
unnormalised singular lenses,
the limit becomes lower ($\Omega_{\Lambda_0} \simeq 0.5$)
with $\sigma_\infty \simeq 1.2 \sigma_{||}$ singular
galaxies.
Even if 
$r_{\rmn c*} = 0.1 R_{\rmn g}$, the upper limit on 
the cosmological constant would still be $\sim 0.6$.
The application of self-consistent dynamics shows that 
the standard models with $\sigma_\infty = \sigma_{||}$ and 
no core radius provide the weakest (\ie\ very 
conservative) upper limits on $\Omega_{\Lambda_0}$.

\section{Conclusions}
\label{section:conclusions}

If the mass distribution of elliptical galaxies is essentially
isothermal, one of the biggest uncertainties in their effect
as gravitational lenses is their dynamical normalisation, $\sigma_\infty$. 
The mass scale depends on their internal dynamics 
(given by $\sigma_{||}$ and $\beta_\sigma$), the luminosity profile
and the aperture used to calibrate the Faber-Jackson (1976)
relation. A range of non-rotating, spherical galaxy models reveals
that $\sigma_\infty \simeq 1.1 \sigma_{||}$ for 
singular models, but that $1 \la \sigma_\infty / \sigma_{||} \la 2$
if ellipticals have significant core radii. 

The dynamical normalisation can have a strong effect on both the
lensing optical depth and the more correct lensing probability. 
In the case of the singular lens model,
both are increased (by up to a factor of 2) by 
the application of the correct normalisation, irrespective of 
the cosmological model. The effect of the normalisation is 
even greater with the presence of a core radius. 
Both the optical depth
and lensing probability increase with very large core radii,
purely due to these self-consistency requirements.
For a given observed Faber-Jackson (1976) relation,
the optical depth can vary by a factor of several with 
both the size of the dispersion aperture and 
the velocity anisotropy.

If lensing statistics are treated primarily as a cosmological 
probe, the above uncertainties place limits on the accuracy of
any cosmological inferences. However, because the correct 
normalisation almost certainly increases the calculated lensing
probability in a given cosmological model, it strengthens 
arguments against a high cosmological constant. 
The weakest limits on $\Omega_{\Lambda_0}$ obtained with the
correct normalisation and an arbitrarily large core radius
are comparable to the limits obtained for the standard
singular models with $\sigma_\infty \simeq \sigma_{||}$.
The dynamical conversions must be 
determined more precisely, theoretically and especially
observationally, to make full use of gravitational 
lensing statistics.

\section*{Acknowledgments}

Many thanks to John Magorrian and Jon Willis for 
stimulating discussions.
DJM was supported by an Australian Postgraduate Award.

\bsp
\label{lastpage}

\begin{thebibliography}{}
\bibitem[\protect\citename{Bahcall \etal\ }1997]{bah97} 
	Bahcall N.\ A., Fan X., Cen R.\ Y., 1997, \ApJ, 485, L53
\bibitem[\protect\citename{Binney \& Tremaine }1987]{bi87}
	Binney J.\ J., Tremaine S., 1987, Galactic Dynamics.
	Princeton University Press, Princeton
\bibitem[\protect\citename{Blandford \& Kochanek }1987]{bl87}
	Blandford R.\ D., Kochanek C.\ S., 1987, \ApJ, 321, 658
\bibitem[\protect\citename{Boyle \etal\ }1999a]{bo99a}
	Boyle B.\ J., Croom S.\ M., Smith R.\ J., Shanks T., Miller L., 
	Loaring N.\ S., 1999a, \PhilTransA, 357, 185
\bibitem[\protect\citename{Boyle \etal\ }1999b]{bo99b}
	Boyle B.\ J., Croom S.\ M., Smith R.\ J., Shanks T.,
        Miller L., Loaring N.\ S., 1999b, 
	in Morganti R., Couch W.\ J., eds,
	Looking Deep in the Southern Sky.\ 
	Springer-Verlag, Berlin, p.\ 16
\bibitem[\protect\citename{Boyle \etal\ }1988]{bo88} 
	Boyle B.\ J., Shanks T., Peterson B.\ A., 1988, \MNRAS, 235, 935
\bibitem[\protect\citename{Carroll \etal\ }1992]{ca92} 
	Carroll S.\ M., Press W.\ H., Turner E.\ L., 1992, \ARAA, 30, 499
\bibitem[\protect\citename{Colless }1999]{co99}
	Colless M.\ M., 1999, 
        in Morganti R., Couch W.\ J., eds,
        Looking Deep in the Southern Sky.\ 
        Springer-Verlag, Berlin, p.\ 9
\bibitem[\protect\citename{de Vaucouleurs }1948]{de48}
	de Vaucouleurs G., 1948, Annalen d'Astrophysics, 11, 247
\bibitem[\protect\citename{de Vaucouleurs \& Olson }1982]{de82}
	de Vaucouleurs G., Olson D.\ W., 1982, \ApJ, 256, 346
\bibitem[\protect\citename{Dressler \etal\ }1987]{dr87}
        Dressler A., Lynden-Bell D., Burnstein D., Davies R.\ L.,
        Faber S.\ M., Terlevich R.\ J., Wegner G., 1987, \ApJ, 
        502, 550
\bibitem[\protect\citename{Dyer \& Roeder }1972]{dy72}
	Dyer C.\ C., Roeder R., 1972, \ApJ, 174, L115
\bibitem[\protect\citename{Dyer \& Roeder }1973]{dy73}
        Dyer C.\ C., Roeder R., 1973, \ApJ, 180, L31
\bibitem[\protect\citename{Efstathiou \etal\ }ef99]{1999}
	Efstathiou G., Bridle S.\ L., Lasenby A.\ N., Hobson M.\ P.,
        Ellis R.\ S., 1999, \MNRAS, 303, L47
\bibitem[\protect\citename{Efstathiou \etal\ } 1988]{ef88} 
	Efstathiou G., Ellis R.\ S., Peterson B.\ A., 1988, \MNRAS, 232, 431
\bibitem[\protect\citename{Faber \& Jackson }1976]{fa76}
	Faber S.\ M., Jackson R.\ E., 1976, \ApJ, 204, 668
\bibitem[\protect\citename{Falco \etal\ }1998]{fa98} 
	Falco E.\ E., Kochanek C.\ S., Mu\~{n}oz J.\ A., 1998,
	\ApJ, 494, 47
\bibitem[\protect\citename{Falco \etal\ }1999]{fa99}
	Falco E.\ E., \etal, 1999, \ApJ, 523, 617
\bibitem[\protect\citename{Folkes \etal\ }1999]{fo99}
	Folkes S.\ R., \etal, 1999, \MNRAS, 308, 459
\bibitem[\protect\citename{Fukugita \& Turner }1991]{fu91}
	Fukugita M., Turner E.\ L., 1991, \MNRAS, 253, 99
\bibitem[\protect\citename{Fukugita \etal\ }1990]{fu90} 
	Fukugita M., Futumase T., Kasai M., 1990,
	\MNRAS, 246, 24
\bibitem[\protect\citename{Gott }1977]{go77}
	Gott J.\ R., 1977, \ARAA, 15, 235
\bibitem[\protect\citename{Guth }1981]{gu81}
	Guth A.\ H., 1981, \PRD, 23, 347
\bibitem[\protect\citename{Hernquist }1990]{he90}
	Hernquist L., 1990, \ApJ, 356, 359
\bibitem[\protect\citename{Hinshaw \& Krauss }1987]{hi87}
	Hinshaw G., Krauss L.\ M., 1987, \ApJ, 320, 468
\bibitem[\protect\citename{Kassiola \& Kovner }1993]{ka93}
        Kassiola A., Kovner I., 1993, \ApJ, 417, 450
\bibitem[\protect\citename{Kayser \etal\ }1997]{ka97} 
	Kayser R., Helbig P., Schramm T., 1997, \AaA, 318, 680
\bibitem[\protect\citename{Keeton \& Kochanek }1996]{ke96}
	Keeton C., Kochanek C.\ S., 1996,
	in Kochanek C.\ S., Hewitt J.\ N., eds,
	Proc.\ IAU Symp.\ No.\ 173.\ Astrophysical Applications of
	Gravitational Lensing.\ Kluwer, Dordrecht, p.\ 419
\bibitem[\protect\citename{Keeton \etal\ }1998]{ke98} 
        Keeton C., Kochanek C.\ S., Falco E.\ E., 1998, \ApJ, 509, 561
\bibitem[\protect\citename{Kochanek }1991]{ko91}
	Kochanek C.\ S., 1991, \ApJ, 379, 517
\bibitem[\protect\citename{Kochanek }1992]{ko92}
	Kochanek C.\ S., 1992, \ApJ, 384, 1 
\bibitem[\protect\citename{Kochanek }1993]{ko93}
	Kochanek C.\ S., 1993, \ApJ, 419, 12
\bibitem[\protect\citename{Kochanek }1994]{ko94}
	Kochanek C.\ S., 1994, \ApJ, 436, 56
\bibitem[\protect\citename{Kochanek }1995]{ko95}
	Kochanek C.\ S., 1995, \ApJ, 453, 545
\bibitem[\protect\citename{Kochanek }1996a]{ko96a}
	Kochanek C.\ S., 1996a, 
        in Kochanek C.\ S., Hewitt J.\ N., eds,
        Proc.\ IAU Symp.\ No.\ 173.\ Astrophysical Applications of
        Gravitational Lensing.\ Kluwer, Dordrecht, p.\ 7
\bibitem[\protect\citename{Kochanek }1996b]{ko96b}
	Kochanek C.\ S., 1996b, \ApJ, 466, 638
\bibitem[\protect\citename{Kochanek \& Blandford }1987]{ko87}
	Kochanek C.\ S., Blandford R.\ D., 1987, \ApJ, 321, 676
\bibitem[\protect\citename{Kochanek \etal\ }1999]{ko99}
	Kochanek C.\ S., Falco E.\ E., Impey C.\ D., Leh\'{a}r J.,
	McLeod B.\ A., Rix H.-W., 1999, 
	in Holt S., Smith E., eds, Am.\ Inst.\ Phys.\ Conf.\
	Proc.\ 470, After the Dark Ages: When Galaxies Were Young
	(The Universe at 2 $ < z < $ 5).\
	Am.\ Inst.\ Phys., Baltimore, p.\ 163
\bibitem[\protect\citename{Kochanek \etal\ }2000]{ko00}
        Kochanek C.\ S., \etal, 2000, \ApJ, in press
\bibitem[\protect\citename{Kolb \& Turner }1989]{ko89}
        Kolb E.\ W., Turner M.\ S., 1989, The Early Universe.\
        Addison-Wesley Publishing Company, Redwood City
\bibitem[\protect\citename{Kormendy \& Djorgovski }1989]{kor89}
	Kormendy J., Djorgovski G., 1989, \ARAA, 27, 235
\bibitem[\protect\citename{Kormendy \& Richstone }1995]{kor95}
	Kormendy J., Richstone D., 1995, \ARAA, 33, 581
\bibitem[\protect\citename{Kormendy \etal\ }1996]{kor96}
	Kormendy J., \etal, 1996, \ApJ, 473, L91
\bibitem[\protect\citename{Kormendy }1997]{kor97}
	Kormendy J., \etal, 1997, \ApJ, 482, L139
\bibitem[\protect\citename{Krauss \& White }1992]{kr92}
	Krauss L.\ M., White M., 1992, \ApJ, 394, 385
\bibitem[\protect\citename{Lauer \etal\ }1995]{la95}
	Lauer T.\ R., \etal, 1995, \AJ, 110, 2622
\bibitem[\protect\citename{Lineweaver }1998]{line98}
	Lineweaver C.\ H., 1998, \ApJ, 505, L69
\bibitem[\protect\citename{Loveday \& Pier }1998]{lov98}
	Loveday J., Pier J., 1998, 
	in Colombi S., Mellier Y., Raban B., eds, 
	Wide Field Surveys in Cosmology.\ 
	Edition Frontiers, Paris, p.\ 317
\bibitem[\protect\citename{Malhotra \etal\ }1996]{mal97} 
	Malhotra S., Rhoads J.\ E., Turner E.\ L., 1997, \MNRAS, 288, 138
\bibitem[\protect\citename{Mao }1991]{ma91}
        Mao S., 1991, \ApJ, 380, 9
\bibitem[\protect\citename{Mao \& Kochanek }1994]{ma94}
        Mao S., Kochanek C.\ S., 1994, \MNRAS, 268, 569
\bibitem[\protect\citename{Maoz \& Rix }1993]{mao93}
	Maoz D., Rix H.-W., 1993, \ApJ, 416, 425
\bibitem[\protect\citename{Mortlock }1999]{mortthes99}
        Mortlock D.\ J., 1999, PhD Thesis, University of Melbourne
\bibitem[\protect\citename{Mortlock \& Webster }2000a]{mort00a}
	Mortlock D.\ J., Webster R.\ L., 2000a,
	\MNRAS, in press
\bibitem[\protect\citename{Mortlock \& Webster }2000b]{mort00b}
        Mortlock D.\ J., Webster R.\ L., 2000b,
        \MNRAS, in press
\bibitem[\protect\citename{Navarro \etal\ }1996]{nav96}
	Navarro J.\ F., Frenk C.\ S., White S.\ D.\ M., 1996, \ApJ, 462, 563
\bibitem[\protect\citename{Navarro \etal\ }1997]{nav97}
	Navarro J.\ F., Frenk C.\ S., White S.\ D.\ M., 1997, \ApJ, 490, 493
\bibitem[\protect\citename{Perlmutter \etal\ }1999]{pe99}
	Perlmutter S., \etal, 1999, \ApJ, 517, 565
\bibitem[\protect\citename{Press \& Gunn }1973]{pr73}
	Press W.\ H., Gunn J.\ E., 1973, \ApJ, 185, 397
\bibitem[\protect\citename{Rix \etal\ }1994]{rix94}
        Rix H.-W., Maoz D., Turner E.\ L., Fukugita M., 
        1994, \ApJ, 435, 49
\bibitem[\protect\citename{Schechter }1976]{sch76}
	Schechter P., 1976, \ApJ, 203, 297
\bibitem[\protect\citename{Schmidt \etal\ }1998]{schm98}
	Schmidt B.\ P., \etal, 1998, \ApJ, 507, 46
\bibitem[\protect\citename{Schneider \etal\ }1992]{sch92} 
	Schneider P., Ehlers J., Falco E.\ E., 1992, 
	Gravitational Lenses.\ Springer-Verlag, Berlin
\bibitem[\protect\citename{Subramanian \& Cowling }1986]{su86}
	Subramanian K., Cowling S.\ A., 1986, \MNRAS, 219, 333
\bibitem[\protect\citename{Szalay }1998]{sza98}
	Szalay A.\ S., 1998, 
	in M\"{u}ller V., Gottl\"{o}ber S., M\"{u}cket J.\ P.,
 	Wambsganss J., eds,
	Large Scale Structure: Tracks and Traces.\
	World Scientific, Singapore, p.\ 97
\bibitem[\protect\citename{Turner }1980]{turn80}
	Turner E.\ L., 1980, \ApJ, 242, L135
\bibitem[\protect\citename{Turner }1990]{tur90}
	Turner E.\ L., 1990, \ApJ, 365, L43
\bibitem[\protect\citename{Turner \etal\ }1984]{tur84} 
	Turner E.\ L., Ostriker J.\ P., Gott J.\ R., 1984, \ApJ, 284, 1
\bibitem[\protect\citename{van der Marel }1991]{van91}
	van der Marel R.\ P., 1991, \MNRAS, 253, 710
\bibitem[\protect\citename{Wallington \& Narayan }1993]{wal93}
	Wallington S., Narayan R., 1993, \ApJ, 403, 517
\bibitem[\protect\citename{Young }1976]{you76}
	Young P.J., 1976, \ApJ, 81, 807 
\end{thebibliography}
\end{document}